\documentclass[useAMS,usenatbib]{mn2e}
\usepackage{graphicx}    
\usepackage{subfig}
\usepackage{amssymb}
 
\DeclareGraphicsExtensions{.ps,.pdf,.png}
\makeatletter
\def\fps@figure{htbp}
\makeatother

\begin{document}

\title[The Growth of Massive Galaxies]{Minor vs Major Mergers: The Stellar Mass Growth of Massive Galaxies from z=3 using Number Density Selection Techniques}
\author[Ownsworth et al.]{Jamie~R.~Ownsworth$^{1}$\thanks{E-mail: ppxjo1@nottingham.ac.uk}, Christopher~J.~Conselice$^1$, Alice~Mortlock$^{1,2}$,  \newauthor
William~G.~Hartley$^{1,3}$, Omar~Almaini$^{1}$, Ken~Duncan$^{1}$, Carl~J.~Mundy$^{1}$ \\
$^{1}$University of Nottingham, School of Physics and Astronomy, Nottingham, NG7 2RD, U.K. \\
$^{2}$SUPA\thanks{Scottish Universities Physics Alliance}, Institute for Astronomy, University of Edinburgh, Royal Observatory, Edinburgh, EH9 3HJ, U.K. \\
$^{3}$Institute for Astronomy, ETH Zurich, Wolfgang-Pauli-Strasse 27, CH-8093 Zurich, Switzerland}
\date{Accepted 2nd Sept 2014. Received 28th Aug 2014; in original form 20th March 2014}
\pagerange{\pageref{firstpage}--\pageref{lastpage}} \pubyear{2014}
\maketitle

\label{firstpage}
\begin{abstract}

We present a study on the stellar mass growth of the progenitors of local massive galaxies with a variety of number density selections with $n\le1\times10^{-4}\,\rm{Mpc^{-3}}$ (corresponding to $M_*=10^{11.24}\rm{M_{\odot}}$ at z=0.3) in the redshift range $0.3<z<3.0$. We select the progenitors of massive galaxies using a constant number density selection, and one which is adjusted to account for major mergers. We find that the  progenitors of massive galaxies grow by a factor of four in total stellar mass over this redshift range. On average the stellar mass added via the processes of star formation, major and minor mergers account for $24\pm8\%$, $17\pm15\%$ and $34\pm14\%$, respectively, of the total galaxy stellar mass at $z=0.3$. Therefore $51\pm20\%$ of the total stellar mass in massive galaxies at $z=0.3$ is created externally to their z=3 progenitors.  We explore the implication of these results on the cold gas accretion rate and size evolution of the progenitors of most massive galaxies over the same redshift range. We find an average gas accretion rate of $\sim66\pm32\,\rm{M_{\odot}yr^{-1}}$ over the redshift range of $1.5<z<3.0$. We find that the size evolution of a galaxy sample selected this way is on average lower than the findings of other investigations.
\end{abstract}

\begin{keywords}
galaxies:evolution, galaxies:high-redshift, galaxies:star formation, galaxies:interactions, galaxies:structure, infrared:galaxies
\end{keywords}

\section{Introduction}
\label{sec:Intro}

The main process by which galaxies acquire their stellar mass and gas is still an open question in galaxy formation. We know from galaxy stellar mass functions that galaxies increase in stellar mass over time (e.g. \citealt{Cole2001}, \citealt{PGon2008}, \citealt{Ilbert2010}, \citealt{Mortlock2011}, \citealt{Muzzin2013}). We also know that there are at least two primary processes via which galaxies can increase their stellar mass; star formation and merging of pre-existing galaxies. However, it has been very difficult to disentangle these two processes primarily as it is challenging to link descendants and progenitors of galaxies at different redshifts.

A common solution for linking galaxies at different redshifts is to examine galaxies at a fixed stellar mass. This is however only truly effective at selecting galaxies that have undergone passive evolution over the examined redshift range, e.g. luminous red galaxies (e.g. \citealt{Wake2006}) assuming there are no mergers.  However the general population of galaxies at high redshift are not passively evolving but show signs of recent large amounts of star formation (e.g. \citealt{Daddi2007}, \citealt{Bauer2011}, \citealt{O2012} , \citealt{Dokkum2013}) and mergers (e.g. \citealt{Conselice2006}, \citealt{Bluck2009}, 2012). 

Recent studies (e.g. \citealt{Dokkum2010}, \citealt{Papovich2011}, \citealt{C2013}, \citealt{Marchesini2014}, \citealt{Lundgren2014}) introduced a new approach to help solve this problem by tracing galaxies at a constant number density. This approach assumes that the relative number density of the most massive galaxies does not evolve i.e. they undergo very few mergers with galaxies of similar stellar mass over the redshift range studied. This technique has been used to examine the evolution of a number of galaxy properties e.g. star formation histories at $z>3$ (\citealt{Papovich2011}, Salmon et al. in prep), as well as structural parameters and stellar mass (\citealt{Dokkum2010}, \citealt{Patel2013}, \citealt{C2013}). Semi\--analytical methods have shown the constant number density selection to be a considerable improvement in tracking the evolution of an individual galaxy population over $0<z<3$ compared to previous mass selection techniques (\citealt{Leja2013}). 

Using a constant number density selection to trace galaxy population however does have its limitations. For example, \cite{Behroozi2013b} and \cite{Leja2013} find that a constant number density selection in semi\--analytical models over the redshift range of $z=0$ to $z=3.0$ could only reproduce the median stellar mass growth of descendants of the most massive galaxies to within 40\% of the ``true'' value in the model. This offset can be reduced to 12\% when this number density is adjusted for the galaxies destroyed via mergers.  In practise however, we are just now starting to measure the merger history with any accuracy.  To make further progress with tracing galaxy populations through time the number density selection must be adjusted at each redshift to account for major mergers that occur within this population.   

Mergers are of course important in themselves, as in the hierarchical picture of galaxy formation massive objects form by the merging together of smaller objects. As such, galaxies will be undergoing mergers at all redshifts. Over redshifts $0<z<3$ close pair and morphological methods find a positive evolution of the major merger fraction with redshift (e.g. \citealt{Bluck2009}, 2012, \citealt{Bridge2010}). From a theoretical perspective, in the $\Lambda$ Cold Dark Matter ($\Lambda$CDM) paradigm dark matter halos form from the bottom up, with larger halos created at later times (e.g. \citealt{Lacy1993}, \citealt{Springel2005}). As galaxies lie inside these haloes they trace the underlying dark matter distribution, and therefore we expect these to undergo hierarchical growth as well. However, it has been shown that some massive galaxies exist and have old stellar populations in place at high redshifts (e.g.\citealt{McCarthy2004}, \citealt{Daddi2005},  \citealt{Bauer2011}, \citealt{Mortlock2011}, \citealt{Hartley2013}). This implies that these galaxies must undergo rapid evolution at early times in the universe, or that some distant mergers are 'dry'.

Galaxy formation is likely driven, at least in part, by mergers. But there are other processes that account for the build up of stellar mass, most especially the star formation rate.  The peak in the volume averaged star formation rate for all galaxies in the Universe occurs in the redshift range of $1.5<z<2.5$ (e.g. \citealt{Madau1996}, \citealt{Hopkins2006}, \citealt{Tresse2007}, \citealt{Wilkins2008}, \citealt{Behroozi2013}). Within this epoch, the star formation rate in typical galaxies is an order of magnitude higher than in the local universe (e.g. \citealt{Reddy2009}). Studies of massive galaxies show a similar trend whereby at high redshift they experience high star formation rates (SFRs) that decrease towards lower redshifts (e.g. \citealt{Daddi2007},  \citealt{Dokkum2010}, \citealt{Bauer2011} \citealt{O2012}). However, the SFRs of the most massive galaxies in the Universe peaks earlier than the total galaxy population at around $z\sim3$ (\citealt{Papovich2011}). This reveals that the galaxy population is experiencing the effects of downsizing, wherein the most massive galaxies shut off their star formation before lower mass objects.

Perhaps related to this, there also exists a tight correlation and a low scatter between SFRs and stellar mass over a large range of redshifts for star forming galaxies (\citealt{Daddi2007}, \citealt{Noeske2007}, \citealt{Pannella2009},  \citealt{Magdis2010}). These studies suggest that massive galaxies at high redshift sustain high levels of star formation for extended amounts of time. The high star formation rates (SFRs) experienced by massive galaxies are fuelled by the large cold gas fraction found in galaxies at high redshift compared to low redshift (e.g. \citealt{Tacconi2010}). The high levels of star formation in massive galaxies would however exhaust these gas reservoirs on very short time scales, $\sim500$Myr (\citealt{C2013}). Therefore it can be inferred that the difference between the integrated SFR and the total stellar mass must correspond to the stellar mass acquired via mergers over $0.3<z<3.0$. 

To understand these issues, and to come up with a coherent picture of galaxy formation, we present a study of the stellar mass growth of the progenitors of local massive galaxies at number densities of $n<1\times10^{-4}\,\rm{Mpc^{-3}}$ in the redshift range $0.3<z<3.0$ by examining all of these formation processes.  We indirectly measure the minor merger rates of the progenitors of local massive galaxies at early cosmic times using a major merger adjusted number density technique. From this we measure the relative contributions of star formation, major and minor merger to the total stellar mass growth of these progenitor galaxies. This will help us understand how and when the most massive galaxies in the universe assembled their stellar mass. 

The paper is set out as follows: \S 2 discusses the Ultra Deep Survey and how the data used in this paper was obtained including the redshifts, stellar masses and star formation rates. \S 3 discusses the galaxy number density selection methods. \S 4.1 presents the results of the stellar mass growth of the progenitors of massive galaxies from $z=3.0$. \S 4.2 presents the star formation history of the progenitors of massive galaxies from the two selection methods. In \S4.3  we calculate the contribution of minor mergers to the total stellar mass growth. \S 4.4 examines the contributions of all stellar mass growth processes over the redshift range of $0.3<z<3.0$. In \S 4.5  we use the results from this paper to examine the implications for the cold gas accretion rate from the intergalactic medium of the progenitors of local massive galaxies. \S 4.6 examines the evolution in the sizes of the progenitors across the whole redshift range. Finally \S 5 summarises our findings. 

Throughout this paper we assume $\Omega_{M}=0.3$, $\Omega_{\lambda}=0.7$ and $H_0=70$ km s$^{-1}$ Mpc$^{-1}$. AB magnitudes and a Chabrier IMF are used throughout.

\section{Data and Analysis}
\label{sec:Data}

\subsection{The UDS}
\label{sec:UDS}
This work is based on the 8th data release (DR8) of the Ultra Deep Survey (UDS; Almaini et al in prep.), which is the deepest of the UKIRT (United Kingdom Infra\--Red Telescope) Infra\--Red Deep Sky Survey (UKIDSS; \cite{Lawrence2007}) projects. The UDS covers 0.77 deg$^{2}$ in J, H, K and the limiting magnitudes (AB), within an aperture of 2 arcsec and at a 5$\sigma$ level, are 24.9, 24.2, 24.6 in J, H, K respectively. It is the deepest infra\--red survey ever undertaken over such an area. It benefits from an array of ancillary multi\--wavelength data: U\--band data from CFHT Megacam (Foucoud et al. in prep); B,V, R, i$^{\prime}$ and z$^{\prime}$ \--band data from the Subaru-XMM Deep Survey (SXDS; \citealt{Furusawa2008}); infrared data from the Spitzer Legacy Program (SpUDS, PI: Dunlop). All of these are fundamental for the computation of accurate photometric redshifts, stellar masses and rest-frame magnitudes. The galaxy catalogue employed in this work is K-band selected and contains approximately 96000 galaxies. This survey reaches a depth of K$_{AB}$=24.4, which was determined from simulations and guarantees a 99\% completeness level. See \cite{Hartley2013} for more details.

The depth and wavelength of the UDS allows us to study the distant Universe with fewer biases against red and dusty galaxies, which could otherwise be completely missed in ultraviolet and optical surveys.

\subsection{Redshifts}
\label{sec:redshifts}

Photometric redshifts are determined by fitting template spectra to photometry from the following bands: U, B, V, R, i$^{\prime}$, z$^{\prime}$, J, H, K, 3.6$\mu$m and 4.5$\mu$m, with a K-band apparent magnitude prior. The package employed for the template fitting was \textsc{eazy} \citep*{EAZY2008}. The template fitting makes use of the standard six \textsc{eazy} templates and an extra one, a combination of the bluest \textsc{eazy} template with a small amount SMC-like extinction \citep{Prevot1984}. Furthermore, $\sim$1500 spectroscopic redshifts from the UDSz programme (an ESO Large Programme; PI Almaini) are also used to train the fitting procedure. Following the comparison to spectroscopic redshifts from the UDSz programme, and $\sim$4000 archival spectroscopic redshifts, and the removal of obvious AGN and catastrophic outliers ($\delta z/(1+z)>0.15$), the dispersion between the photometric and the spectroscopic redshifts is measured as $\delta z/(1+z)\sim0.031$ (\citealt{Hartley2013}). 

\subsection{Stellar Masses \& SED fitting}
\label{sec:mass}
The stellar masses and rest\--frame colours of our sample are measured using a multicolour stellar population fitting technique. For a full description see \cite{Mortlock2013} and \cite{Hartley2013}. We fit synthetic spectral energy distributions (SEDs) constructed from the stellar populations models of \cite{Bruzual2003} to the U, B, V, R, i$^{\prime}$, z$^{\prime}$, J, H, K  bands and IRAC Channels 1 and 2, assuming a Chabrier initial mass function. The star formation history is characterised by an exponentially declining model with various ages, metallicity and dust content of the form
\begin{equation}
SFR(t)= SFR_{0} \times \rm{exp}(-t/\tau)  
\end{equation}
where the values of $\tau$ ranges between 0.01 and 13.7 Gyr, the age of the onset of star formation ranges from 0.001 to 13.7 Gyr. We exclude templates that are older than the age of the Universe at the redshift of the galaxy being fit. The metallicity ranges from 0.0001 to 0.1 solar, and the dust content is parametrised, following \cite{Charlot2000}, by $\tau_{v}$, the effective V\--band optical depth. We use values up to $\tau_{v}=2.5$ with a constant inter\--stellar medium fraction of 0.3. 
To fit the SEDs they are first scaled in the observed frame to the K\--band magnitude of the galaxy. We then fit each scaled model template in the grid of SEDs to the measured photometry of each individual galaxy. We calculate $\chi^{2}$ values for each template, and select the best fitting template, obtaining a corresponding stellar mass and rest\--frame luminosities. 
\cite{Hartley2013}, following the method from \cite{Pozzetti2010}, found the $95\%$ mass completeness limit  of $M_{\rm{lim}}=8.27+0.81z-0.07z^2$. Galaxies that fall below $M_{\rm{lim}}$ are not used in the subsequent analysis.

\subsection{Galaxy Structural Parameters}
We calculate structural parameters measured on ground based UDS K\--band images using \textsc{galapagos} (Galaxy Analysis over Large Area: Parameter Assessment by \textsc{galfit}ing Objects from SE\textsc{xtractor}; \citealt{Barden2012}). This program uses SE\textsc{xtractor} and \textsc{galfit} to fit S\'{e}rsic light profiles (\citealt{sersic1968}) to objects in the UDS field. An S\'{e}rsic light profile is given by the following equation:
\begin{equation}
\Sigma(R)=\Sigma_{e}\times \rm{exp}\left(-b_{n}\left[\left(\frac{R}{R_{e}}\right)^{1/n}-1\right]\right)
\end{equation}
\noindent Where $\Sigma(R)$ is the surface brightness as a function of the radius, $R$; $\Sigma_{e}$ is the surface brightness at the effective radius, $R_{e}$; $n$ is the S\'{e}rsic index and $b_{n}$ is a function dependant on the S\'{e}rsic index. The sizes (effective radius) are calibrated with galaxy sizes derived from the UDS area from the Hubble Space Telescope (HST) Cosmic Assembly Near\--infrared Deep Extragalactic Legacy Survey (CANDELS) (\citealt{Grogin2011}, \citealt{Koekemoer2011}) by \cite{Wel2012}. For a full description of this method see \cite{Lani2013}. \cite{Lani2013} show that the ground based size measurements are reliable for galaxies with $K < 22$ in the UDS. In Sections 4.5 and 4.6 galaxies that fall below $K < 22$ are not used in the subsequent analysis.

\subsection{Star Formation Rates}
\label{sec:SFRexplan}
We determine the star formation rates within galaxies over the redshift range $0.3 < z < 3$.
Determining the star formation activity at these redshifts is however not trivial. Infra-red observations are useful indicators of dust heating due to star formation, but the Spitzer Space Telescope observations are not deep enough to accurately detect a full mass selected sample of galaxies as only a small number ($\sim10\%$) of the whole sample are detected at $24\mu m$ above a flux limit of $300\mu$Jy (\citealt{C2013}, \citealt{Hilton2012}). 

The SED fitting procedure described in \S \ref{sec:mass} also cannot be used to retrieve a value for the $24\mu m$ flux for our sample due to the lack of photometric data points in this part of the spectrum. However the photometric bands used in the SED fitting correspond to the rest-frame UV and optical wavelengths over the redshift range of this survey and therefore this part of the spectrum is well constrained. This enables us to use the dust corrected rest frame UV as an indicator of the star formation rate of these galaxies. 

\subsubsection{UV SFRs}

The rest-frame UV light traces the presence of young and short-lived stellar populations produced by recent star formation. The star formation rates can be calculated from scaling factors applied to the luminosities. These scaling factors are dependent on the assumed IMF (\citealt{Kennicutt1983}). However, UV light is very susceptible to dust extinction and a careful dust correction has to be applied. The correction we use here is based on the rest frame UV slope as explained in the following section.

The raw 2800\AA\, NUV star formation rates ($SFR_{2800,\rm{SED}}$) used in this paper are obtained from the rest\--frame near UV luminosities measured from the best fit SED model found in the stellar mass fitting. We determine the dust\--uncorrected SFRs, $SFR_{2800,\rm{SED,uncorr}}$, for $z = 0.5\--3$ galaxies from applying the \emph{Galaxy Evolution Explorer} (GALEX) NUV filter to the best fit individual galaxy SED.

To measure the SFR  we first derive the UV luminosity of the galaxies in our sample, then use the \cite{Kennicutt1998} conversion from 2800\AA\, luminosity to SFR assuming a Chabrier IMF:

\begin{equation}
SFR_{UV} (\mathrm{M_{\odot}yr^{-1}}) = 8.24 \times 10^{-29} L_{2800} (\mathrm{ergs\,\, s^{-1}\, Hz^{-1})} 
\end{equation}
This however does not account for dust obscuration which can significantly influence the measured SFR.

\subsubsection{Dust Corrections}

To obtain reliable star formation rates in the rest-frame ultraviolet, we need to account for the obscuration due to dust along the line of sight. \cite{Meurer1999} found a correlation between attenuation due to dust and the rest-frame UV slope, $\beta$, for a sample of local starburst galaxies
\begin{equation}
f_{\lambda} \sim \lambda^{\beta}
\end{equation}

\noindent where $f_{\lambda}$ is the flux density per wavelength interval and $\lambda$ is the central rest wavelength. Using the ten UV windows defined by \cite{Calzetti1994} we measure $\beta$ values from the best fitting SED template. This can be done as the redshift range we examine has well calibrated UV SED fits due to many of the input photometric bands lying in the UV part of the spectrum.
This $\beta$ value is then converted to a UV dust correction using the \cite{Fischera2005} (FD05) dust model. 

However, recent work by \cite{Wijesinghe2010,Wijesinghe2012} on local galaxies using the GALEX probe has shown that a FD05 dust model with the 2200\AA\, feature removed is a better correction to the general population of galaxies than the \cite{Calzetti2001} dust model, which is mainly applied to only highly  star forming systems. We note that at the wavelength range we examine in this paper there is very little difference in the dust correction given by the two models.

Using the \cite{Meurer1999} description of the attenuation, and converting it to attenuation at 2800\AA\, using the FD05 dust model, we derive the equation:
 \begin{equation}
A_{2800}=1.67\beta+3.71
\end{equation}

\noindent  One caveat in correcting for the dust extinction in this way is that the $\beta$ parameter is also effected by the age of the stellar population. A galaxy with an old and passive stellar population will, in the UV part of the spectrum, look very similar to a very highly dust extincted young and star forming galaxy population. This is a problem that can cause massive galaxies to artificially appear to have a very high dust content and thus high star formation rates. 

This problem can be corrected via selecting out the galaxies that are passive via other methods. For theses galaxies we can assume the $\beta$ parameter will be driven by the old stellar populations, not dust attenuation. The selection we use is based on the U, V and J Bessel band rest frame luminosities. These were used by \cite{Williams2009} to select evolved stellar populations from those with recent star formation at $z < 2$. This technique is also used in \cite{Hartley2013} to extend the passive galaxy selection out to higher redshifts. The selection criteria for passive galaxies are as follows:

 \begin{equation}
U-V > 0.88 \times V-J+0.69	(z<0.5)
\end{equation}
\begin{equation}
U-V > 0.88 \times V-J+0.59	(0.5<z<1.0)
\end{equation}
\begin{equation}
U-V > 0.88 \times V-J+0.49	(z>1.0) 
\end{equation}
with $U-V >1.3$ and $V-J < 1.6$ in all cases. The objects that are selected via this method are assigned to a passive category of galaxies. The dust correction derived from the $\beta$ parameter therefore is not used when calculating the SFR for these systems. 

To determine the dust content of passive galaxies we refer to recent studies from the Herschel space mission. \cite{Bourne2012} show from stacking that star forming and passive galaxies have similar dust masses. This possibly indicates that both populations have a similar average UV dust correction. Therefore within a given redshift bin we use the average dust attenuation from star forming galaxies with similar stellar masses as the dust attenuation for passive galaxies. However if we assume these galaxies contain no dust and therefore require no dust correction, then the star formation rates for the passive galaxies are on average a factor of $\sim3$ lower than the average dust corrected star formation rates. The effect of changing the dust correction are discussed in sections 4.2, 4.3 and 4.4, but this does not significantly effect the conclusions of this paper. The true dust correction may lie between these two corrections we apply here, implying that the two sets of SFRs for passive galaxies we present are upper and lower bounds.

Although these criteria efficiently select galaxies with old stellar populations, there is a possibility that the sample could still be contaminated by dusty star forming galaxies, edge on disks or AGN. We minimise this contamination by using the wealth of multi-wavelength data that is available in the UDS field. We cross match our sample with surveys on the UDS field taken at X-ray and radio wavelengths. 

For the X-ray we use data from the Subaru/XMM-Newton Deep Survey (\citealt{Ueda2008}) which covers the UDS field over the energy range of 0.5 keV to 10 keV. For the radio we use \cite{Simpson2006} which utilises VLA 1.4 GHz data. We remove any galaxies that have either a detection in the X-ray or radio to clean this sample of AGN. This data will only effectively select out AGN at $z \lesssim 1$ due to the limits of these surveys, and will only be able to select the most radio loud and very active AGN at higher redshifts.

Furthermore the $24\rm{\mu m}$ data from the SpUDS provides a way to identify red objects that harbour dust-enshrouded star formation. Therefore any objects with a $24\rm{\mu m}$ detection  ($300\rm{\mu}$Jy, $15\sigma$) are assumed to be dusty star forming objects. Any galaxy shown to be passive via the UVJ selection criteria, but which has a $24\rm{\mu m}$ source associated with it will be reassigned to the star forming population and have a full UV dust correction applied. In total $\sim2\%$ of objects selected via the UVJ criteria were reassigned to the star\--forming sample through this method.

\begin{figure*}
\includegraphics[scale=1.0]{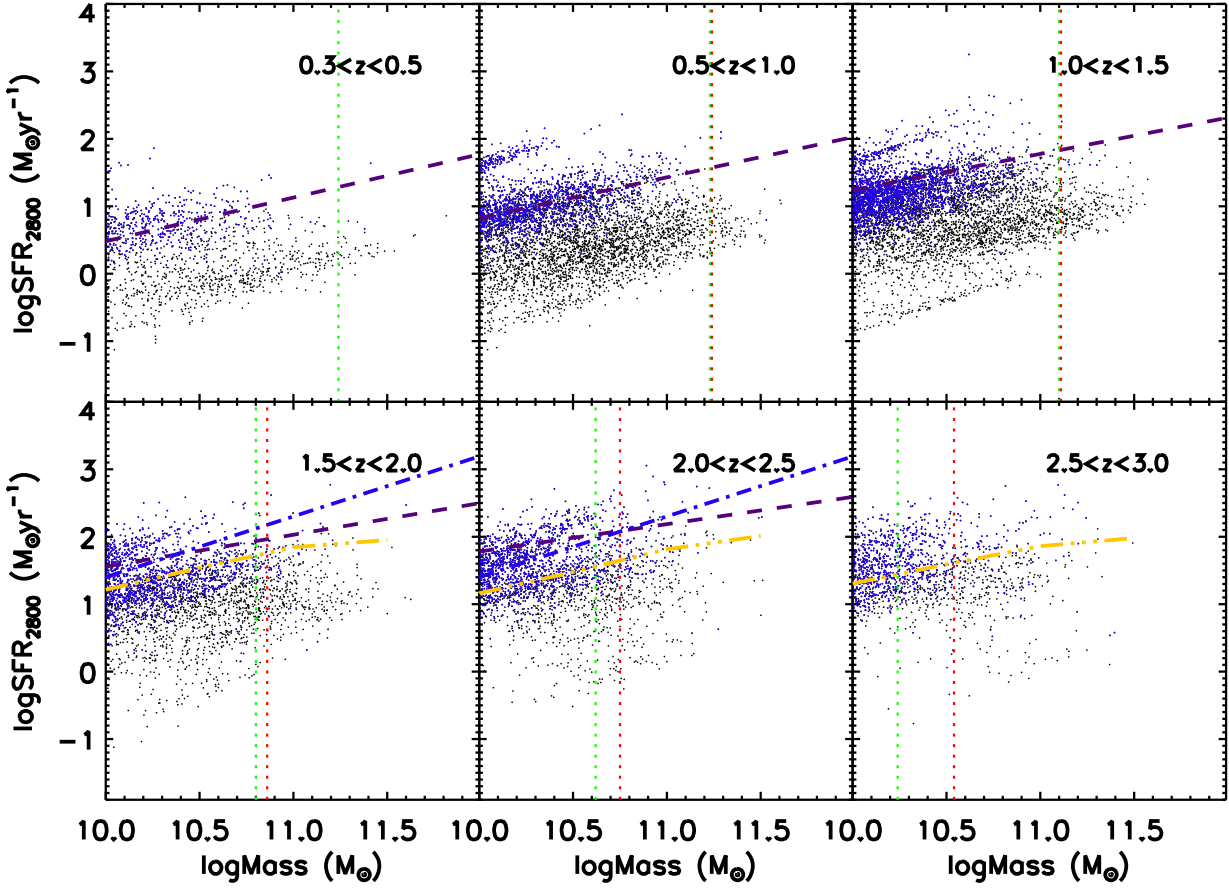}           
\caption{The dust corrected UV star formation rates for all galaxies in the UDS sample as a function of stellar mass. The black points show individual galaxies in the total UDS galaxy catalogue that have been classified as passive using the UVJ criteria described in \S 2.5.2. The blue points show individual  star forming galaxies in the UDS galaxy catalogue. The red and green dotted vertical lines show the stellar mass limits given in Table 1 and 3 denoting the stellar mass limits of the constant number density (red) and major merger adjusted number density (green) selections. The blue dot dashed line is the relation found in Daddi et al. (2007) denoting the relation between the total stellar mass and star formation rate for star forming galaxies between $1.4<z<2.5$. The purple dashed line is the SFR stellar mass relation from Whitaker et al. (2012) using IR+UV SFRS. The yellow treble dot dashed line is the SFR stellar mass relation from Bauer et al. (2011).}
\label{sfrvsmass} 
\end{figure*}

Figure \ref{sfrvsmass} shows SFR versus the stellar mass for all galaxies in the UDS galaxy sample separated into redshift bins. The black points show galaxies that have been classified as passive via the UVJ selection criteria, and blue points show the remaining star forming galaxies. The dotted lines show the stellar mass limits corresponding to the number density selection described in the following section derived from the integrated stellar mass functions of the different galaxy selections. The dashed lines show relations between the SFR and stellar mass of star forming galaxies found by \cite{Daddi2007} at $1.4<z<2.5$, \cite{Whitaker2012} at $0<z<2.5$ and \cite{Bauer2011} at $1.5<z<3.0$. Our $\rm{SFR}_{2800}$  are in good agreement with these relations.

\section{Sample Selection}

In this study we use two selection methods, a constant and a merger adjusted galaxy number density selection. The constant galaxy number density selection uses the number density of the most massive galaxies in the local universe to select the direct progenitors of the most massive galaxies at higher redshifts. The merger adjusted galaxy number density selection is a relatively new method that incorporates in the measured major merger rate of massive galaxies over the redshift range studied. This method selects all of the progenitors of the most massive galaxies, and all major merger progenitor galaxies. This selection method allows us to disentangle between the stellar mass growth of major and minor mergers. In the following sections we describe these two selection methods.

\label{sec:Sample}
\subsection{Constant Galaxy Number Density (C\--GaND)}
A few studies to date have examined galaxy formation and evolution using galaxy number density as a method of selecting galaxies over a large redshift range (e.g. \citealt{Dokkum2010}, \citealt{Papovich2011}, \citealt{C2013}). Several studies have shown that this method of selecting galaxies has several advantages. In the absence of major mergers, or extreme changes of star formation, the number density of galaxies above a given density threshold is invariant with time. These galaxies will grow in stellar mass through star formation and minor mergers, but their number density will stay constant.

In principle, selecting galaxies at a constant number density directly tracks the progenitors and descendants of massive galaxies at all redshifts. A study by \cite{Leja2013} showed that this technique is robust at linking descendant and progenitor galaxies over cosmic time when compared to semi\--analytic models that trace individual galaxies evolving over the last eleven billion years.

In this study we select and compare galaxies at constant co-moving number density values of $n=5\times10^{-4}\, \rm{Mpc^{-3}}$, $n=1\times10^{-4}\, \rm{Mpc^{-3}}$, and $n=0.4\times10^{-4}\, \rm{Mpc^{-3}}$ at redshifts $0.3<z<3$. We chose these number densities as a trade\--off between having a robust number of galaxies in the analysis at each redshift, and retaining a mass complete sample at the highest redshifts. This number density range is comparable to number densities used in other similar studies (e.g. \citealt{Papovich2011}, \citealt{C2013}). 

\begin{figure*}
\subfloat[]{\includegraphics[scale=0.5]{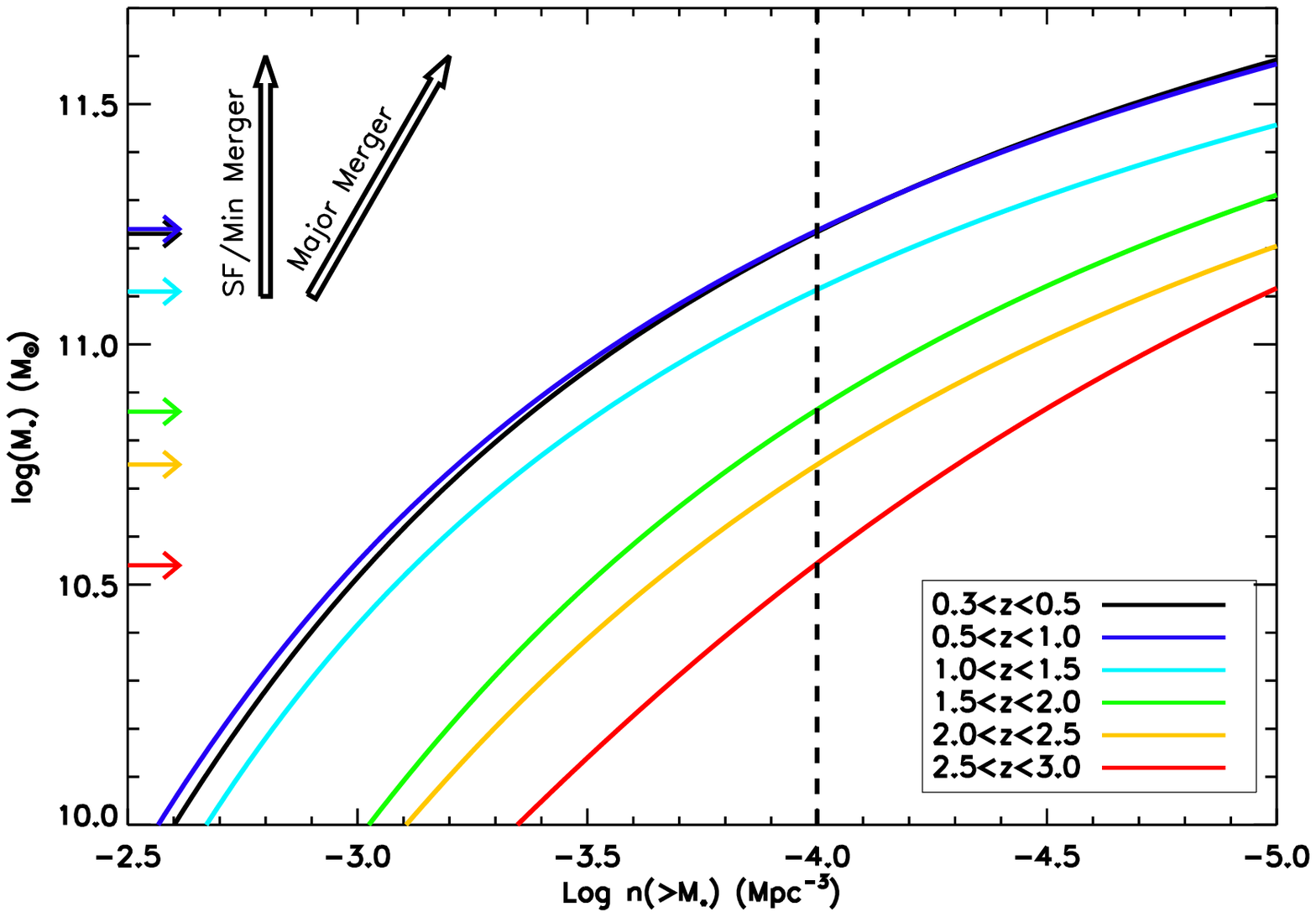}}
\subfloat[]{\includegraphics[scale=0.5]{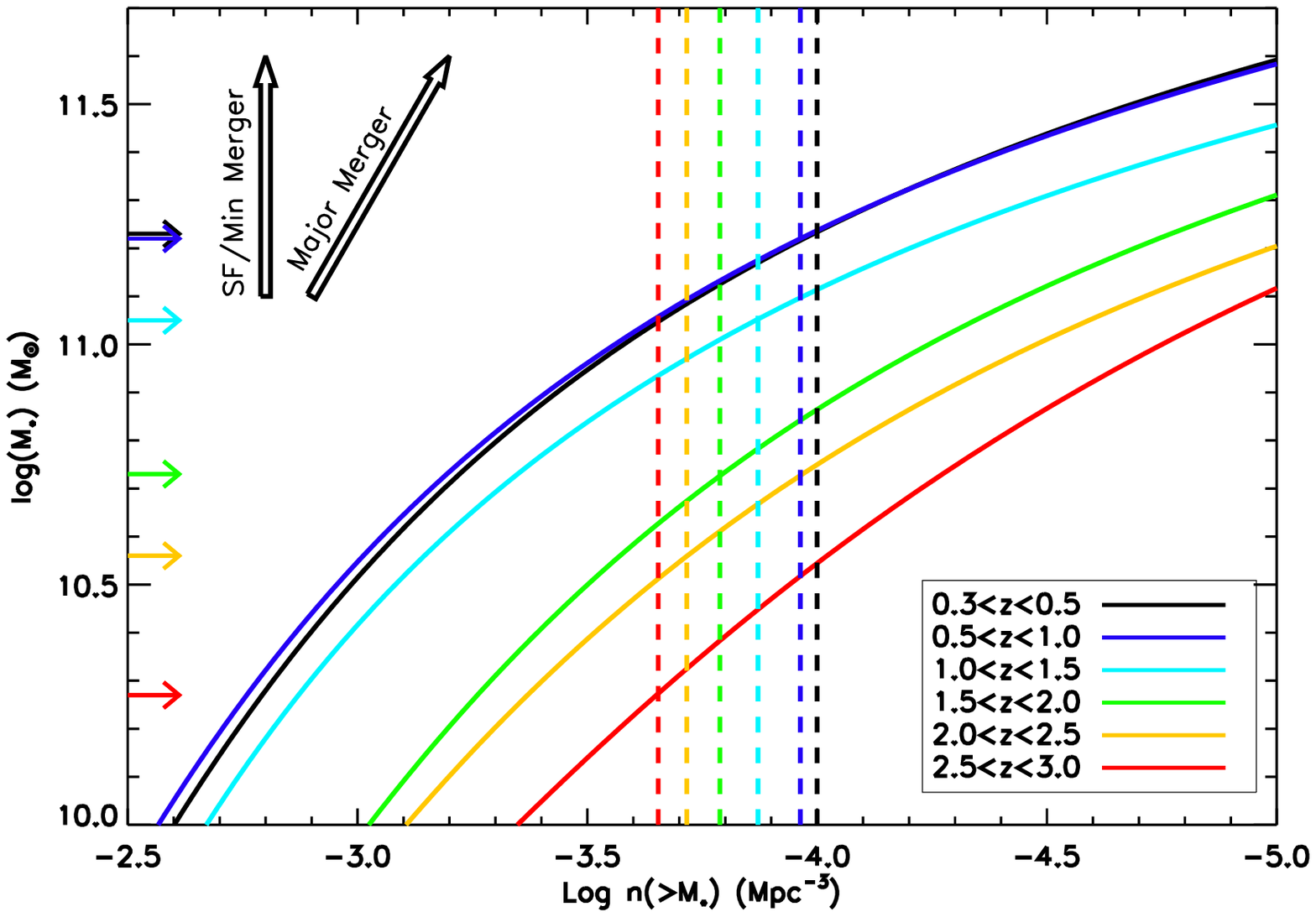}}
\caption{The integrated stellar mass functions from $z=0.3$ to $z=3$ from Mortlock et al. (2014, in prep). These integrated stellar mass functions gives us the co-moving number density of all galaxies more massive than a given stellar mass. The large open black arrows indicate the expected evolution due to star formation, minor mergers and major mergers. (a) We compare galaxies at a constant number density by selecting galaxies at each redshift at limits of $n(>M_{*}) = 1\times 10^{-4}\rm{Mpc^{-3}}$. The black dashed vertical line denotes the constant number density of $1\times10^{-4}\rm{Mpc^{-3}}$. The coloured arrows indicate the values of $M_{*}$ that correspond to this number density for each integrated stellar mass fraction. (b) The galaxy selection using an evolving number density based on the major merger rate from Bluck et al. (2012). by selecting galaxies at each redshift such that $n(>M_{*})$ equals the values for each redshift given in Table 3. The coloured dashed lines denote the number density selection for each redshift. The coloured arrows indicate the values $M_{*}$ that correspond to this number density for each integrated stellar mass function. }
\label{smf}
\end{figure*}

\begin{figure}
\includegraphics[scale=0.5]{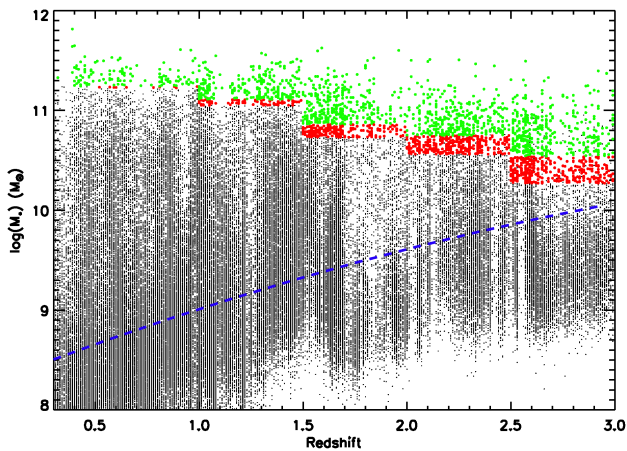}           
\caption{Stellar mass versus photometric redshift for the UDS galaxy parent sample. The blue dashed line is a second order polynomial fit to the $95\%$ mass completeness limit at that redshift (Hartley et al. 2013). The green points indicate the galaxies selected via the constant number density selection, and the red and green points combined show the galaxies statistically selected via the evolving number density selection.}
\label{masscomp}
\end{figure}

We select our sample based on the integrated mass functions of the UDS field over the redshift range of $z= 0.3$ to 3.0 from Mortlock et al (2014, in prep). Table 1 shows the Schechter function fitted parameters. Figure \ref{smf} (a) shows the integrated mass functions from Mortlock et al. (2014, in prep) and the lower stellar mass limits for the constant number density selection. The values for the limits are listed in Table \ref{tab:smfc}. The arrows in the top left hand of Figure \ref{smf} show how the galaxy stellar mass functions will change due to the two processes of stellar mass growth explored in this paper. Figure \ref{masscomp} shows, in green, the galaxies selected via this selection compared to the whole galaxy sample over the redshift range in this study.

\begin{center}
\begin{table}
\caption{Stellar mass function Schechter function fitted parameters from Mortlock et al (2014, in prep).}
\begin{center}
\begin{tabular}{| c | c | c | c |}
  \hline
  \hline
  $z$ & $\rm{log(M_{*})} \rm{(M_{\odot})}$ & $\Phi^{*} (\times 10^{-4})$ & $\alpha$ \\
  \hline             
  $0.3-0.5$ & $11.2\pm0.1$ &$ 7\pm3$ &$ -1.4\pm0.1$\\
  $0.5-1.0$ & $11.1\pm0.1$ & $8\pm3$ & $-1.3\pm0.1$\\
  $1.0-1.5$ & $11.0\pm0.1$ &$ 8\pm2$ & $-1.3\pm0.1$\\
  $1.5-2.0$ & $11.0\pm0.1$ & $2\pm2$ & $-1.5\pm0.2$\\
  $2.0-2.5$ & $11.0\pm0.1$ & $2\pm2$ & $-1.5\pm0.2$\\
  $2.5-3.0$ & $11.1\pm0.4$ &$ 1\pm1$ &$ -1.8\pm0.2$\\
  \hline  
\end{tabular}
\label{tab:sfmevo}
\end{center}
\end{table}
\end{center}

\begin{center}
\begin{table}
\caption{C\--GaND stellar mass limits for a constant number density selected sample taken from the integrated mass functions shown in Figure 2 from Mortlock et al (2014, in prep).}
\begin{center}
\begin{tabular}{| c | c | c | }
  \hline
  \hline
  $z$ & $\rm{log}$ $\rm{n(<M_{\odot})} \rm{(Mpc^{-3})}$ & Stellar Mass limit (log$\rm{M_{\odot}}$) \\
  \hline             
  $0.3-0.5$ & -4.00 & $11.24\pm0.07$\\
  $0.5-1.0$ & -4.00 & $11.24\pm0.04$\\
  $1.0-1.5$ & -4.00 & $11.11\pm0.04$\\
  $1.5-2.0$ & -4.00 & $10.86\pm0.05$\\
  $2.0-2.5$ & -4.00 & $10.75\pm0.07$\\
  $2.5-3.0$ & -4.00 & $10.54\pm0.09$\\
  \hline  
\end{tabular}
\label{tab:smfc}
\end{center}
\end{table}
\end{center}

\subsection{Merger Adjusted Galaxy Number Density (M\--GaND)}

\begin{figure}
\includegraphics[scale=0.5]{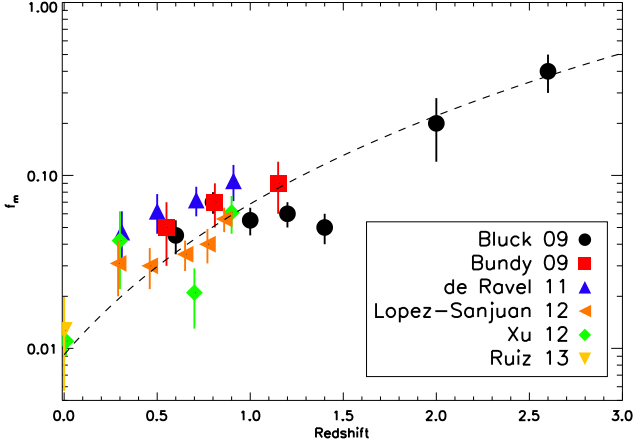}           
\caption{Observed galaxy pair fractions in the literature. \protect \cite{Bluck2009} calculate the merger fraction down to a stellar mass ratio of 1:4 for galaxies with $\rm{log(M_{*})}>11.0$ using close pairs within 30kpc. \protect \cite{Bundy2009} calculate the merger fraction down to a stellar mass ratio of 1:4 for galaxies with $\rm{log(M_{*})}>11.0$ using close pairs within 20kpc. \protect \cite{Ravel2011} calculate the merger fraction of galaxies with $\rm{log(M_{*})}>11.0$ using close pairs within 30kpc and $\Delta B <1.5$. \protect \cite{Lopez2012} calculate the merger fraction down to a stellar mass ratio of 1:4 for galaxies with $\rm{log(M_{*})}>11$ using close pairs within 30kpc. \protect \cite{Xu2012} calculate the merger fraction down to a stellar mass ratio of 1:3 of galaxies with $\rm{log(M_{*})}>10.6$ using close pairs within 20kpc. \protect \cite{Ruiz2013} calculate the merger fraction down to a stellar mass ratio of 1:5 for galaxies with $\rm{log(M_{*})}>11.3$ using close pairs within 100kpc. The \protect \cite{Ruiz2013} point has been modified to compensate for the large close pair search radius. The dashed line is the best fit to all points with the form $\rm{f_{m}}=A\times(1+z)^B$ with $A=0.009\pm0.002$ and $B=2.9\pm0.2$. }
\label{fm}
\end{figure}

Many studies to date have investigated the average number of major mergers a massive galaxies experiences over cosmic time (e.g. \citealt{Bluck2009}, \citealt{Bundy2009}, \citealt{Ravel2011}, \citealt{Lopez2012}, \citealt{Xu2012}, \citealt{Ruiz2013}). Figure \ref{fm} shows the observed pair fractions in the literature which have investigated the major merger rates of massive galaxies using similar methods. Using these merger fractions we can adjust the number density 
 selection to study the contribution of major mergers to the total stellar mass growth. Using both the C\--GaND and M\--GaND selections we can separate the stellar mass growth due to major mergers, star formation and indirectly minor mergers from the total stellar mass growth. We do this using a number density selection that changes due to the rate of major mergers that are occurring between redshift bins. From the best fitting power law to the data shown in Figure \ref{fm} we quantify the fraction of merger events as:

  \begin{equation}
f_{merger}=0.009\pm0.002(1+z)^{2.9\pm0.2} 
\end{equation}
Where $f_{\rm{merger}}$ is the fraction of major merger events at redshift $z$. This relation is derived using galaxies with stellar masses greater than $\rm{log(M)}>11.0$ at all redshifts. \cite{Bluck2012} show that the merger fraction relation with redshift does not change over the stellar mass range of interest in this paper.  

In previous works the merger faction has been converted into a galaxy merger faction, $f_{gm}$. This is appropriate when examining the merger rates within a population. Using \cite{Mortlock2011} we calculate which galaxies below the C\--GaND stellar mass limits are large enough to constitute a 1:4 stellar mass merger ratio.  We find that the number of galaxies below this limit is five times more numerous than galaxies larger than the C\--GaND stellar mass limits. Thus we calculate the number of mergers using $f_{m}$.

From this we calculate the average time between mergers that a galaxy experiences at a given redshift, $\Gamma$, as:
  \begin{equation}
\Gamma = \tau_{\rm{m}}/f_{\rm{m}}
\end{equation}
We adopt a time\--scale over which merging is occurring for galaxy close pairs in a 1:4 or less mass ratio of $\tau_{\rm{m}}=0.4\pm0.2$ Gyr derived from simulation results of \cite{Lotz2008}. We use the $\Gamma$ value to calculate the average number of mergers between our redshift bins using the equation:
  \begin{equation}
N_{\rm{m}}=\int_{t_{1}}^{t_{2}}\frac{dt}{\Gamma(z)}=\int_{z_{1}}^{z_{2}}\frac{1}{\Gamma(z)}\frac{t_{H}}{(1+z)}\frac{dz}{E(z)}
\label{N}
\end{equation}
Where $\Gamma(z)$ is the average time between major mergers, $t_{H}$ is the Hubble time and $E(z)=[\Omega_{\rm{M}}(1+z)^{3}+\Omega_{k}(1+z)^{2}+\Omega_{\Lambda}]^{-1/2}=H(z)^{-1}$. Calculating this from $z=3.0$ to $z=0.3$ we obtain $N_{\rm{m}}=1.2\pm0.5$ as the average number of major mergers that the galaxies selected via the C\--GaND selection will undergo.

Using Equation \ref{N} we calculate the average number of major mergers in each redshift bin. We then compute the major merger adjusted number density via the equation:
  \begin{equation}
n_{z(1)}=n_{z(0)}*(1.0+N_{\rm{m,z(0-1)}})
\end{equation}
\begin{figure}

\hspace{-5mm}\includegraphics[scale=0.5]{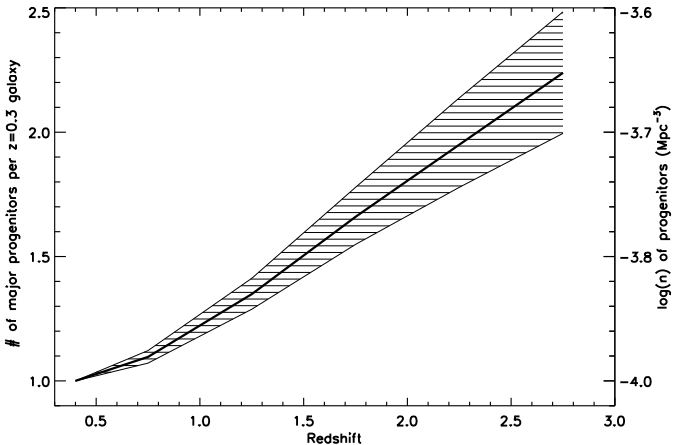}
 
\caption{The mean number of major merger progenitor galaxies against redshift for galaxies with $n=1\times10^{-4}\rm{Mpc^{-3}}$ at $z=0.3$. The solid black line is derived from equation 12. The black hashed area shows the 1 sigma uncertainty on this relation. The y axis on the right hand side shows how the number density of the major merger progenitors evolve with this relation.}
\label{npro}
\end{figure}
\noindent Where $n_{z(0)}$ is the number density of the massive galaxies at redshift $z(0)$. The value $n_{z(1)}$ is the number density of the  progenitors of the galaxies at redshift $z(0)$ at $z(1)$, where $z(1)>z(0)$. $N_{\rm{m,z(0-1)}}$ is the average number of major mergers the progenitor galaxies will experience between $z(1)$ and $z(0)$.
Using this we find that the number density of all the major merger progenitors of local massive galaxies increases with look\--back time by a factor of $2.2$ by redshift $z=3.0$. The exact values of the evolving number densities can be found in Table 3. Figure \ref{smf} (b) shows the integrated galaxy mass functions and lower limit stellar mass cuts based on the evolving number density. Figure \ref{masscomp} furthermore plots the galaxies selected via this method in green and red compared to the total UDS galaxy population. Figure \ref{npro} shows the mean number of progenitor galaxies at each redshift.

Using a major merger adjusted number density selection method we in theory obtain close to a complete sample of the direct progenitors of local massive galaxies, including the less massive galaxies that have merged during a major merger event with the direct central progenitors over the redshift range $0.3<z<3.0$. This selection method also allows us to examine and disentangle the contributions to the total stellar mass growth from major and minor mergers. We achieve this by examining how the stellar mass density of the M\--GaND sample evolves with redshift compared to the C\--GaND sample. The stellar mass density of the M\--GaND sample contains both the stellar mass of the progenitors of local massive galaxies and the stellar mass of the total major merger progenitors. When examining other properties of massive galaxies, e.g. size, across a large redshift range methods that select only the direct progenitors of the local massive galaxies are appropriate.

\begin{figure}
\includegraphics[scale=0.5]{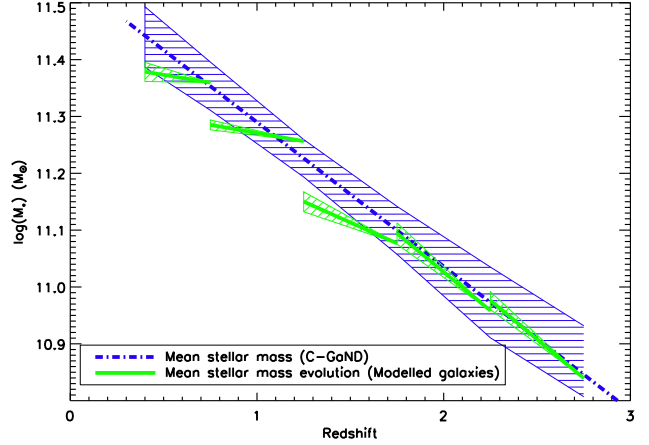}           
\caption{The mean stellar mass evolution of the modelled galaxies. Figure showing how well star formation and major mergers within a given galaxy population is able to account for the change in stellar mass. The blue dot dash line shows the best fit to the evolution of the mean stellar mass of the C\--GaND selected sample with $n=1\times10^{-4}\rm{Mpc^{-3}}$. The blue hashed region shows the 1 sigma uncertainty on this relation. See Section 4.1 for more details.  The green solid lines show the evolution of the mean stellar mass of the galaxies with modelled stellar mass growth. The green hashed regions show the standard error on the mean of these results. The stellar mass growth modelling is described in Section 3.2.}
\label{simgals}
\end{figure}

\subsection{Limits to the Method}

One caveat of selecting galaxies using cuts in stellar mass is contamination from lower mass galaxies entering the sample at lower redshifts or galaxies dropping out due to quenching. This arises due to galaxies below the stellar mass selection limit growing in stellar mass between redshift bins via star formation and mergers. We model this contamination using our knowledge of star formation rates and major merger rates. The stellar mass of each individual galaxy is evolved to the next lowest bin by modelling the star formation histories and major mergers. The stellar mass added via star formation is modelled by integrating the fitted declining $\tau$ model derived from SED fitting for each individual galaxy. The stellar mass added via major mergers is modelled by assigning each galaxy a probability that it will undergo a major merger between redshift bins with a random merger ratio between 1:1 and 1:4. The probability of a major merger is then converted to a number of merger events within a redshift bin by using a Monte Carlo technique. 

Adding together these two stellar mass evolution processes we calculate the evolved stellar mass for each galaxy. We do not take into account the effect of minor merger as we do not fully understand the full influence these events have on the stellar mass growth. Figure \ref{simgals} shows how the mean stellar mass of the galaxies we evolve compares to the evolution of the C\--GaND sample. We find that at high redshifts the modelling appears to more accurately trace the stellar mass evolution of the C\--GaND population than at lower redshifts. This could be due to a higher importance of minor mergers at lower redshifts. From this modelling we find that the number density selection techniques used here has between a $20-30\%$ contamination rate per redshift bin. However the contamination is, on average, three times lower than a constant mass selection technique. We also note that the galaxies with the highest probability of contaminating the sample arise from galaxies within 0.15 dex below the stellar mass limits.
   
When using a merger adjusted number density selection, the exact stellar mass of the smaller galaxy within a major merger is unknown as it could be any galaxy within the mass ratio of 1:4. The selection we use here to construct the M\--GaND sample provides a hard upper limit on the amount of stellar mass that can be assembled via major mergers. This is because we select the most massive galaxies that fall below the C\--GaND selection limit at each redshift. However, constructing the M\--GaND sample this way does result in an apparently sequential merger process i.e., less massive satellites merge first. This is counter to recent findings (e.g. \citealt{Lopez2012}, \citealt{Xu2012}). The stellar mass accretion rates calculated by this work are derived from the total stellar mass densities of both samples, the exact sequence of mergers therefore does not affect the results.

\begin{center}
\begin{table}
\caption{M\--GaND stellar mass limits for the evolving number density sample taken from the integrated mass functions shown in Figure 2 from Mortlock et al (2014, in prep). Starting at log(n)$=-4.0$ in the $z=0.3-0.5$ redshift bin.}
\begin{center}
\begin{tabular}{| c | c | c | }
  \hline
  \hline
  $z$ & $\rm{log}$ $\rm{n(<M_{\odot})}$ & Stellar Mass limit (log$ \rm{M_{\odot}}$) \\
  \hline             
  $0.3-0.5$ & $-4.00$ & $11.24\pm0.07$\\
  $0.5-1.0$ & $-3.96\pm0.01$ & $11.22\pm0.04$\\
  $1.0-1.5$ & $-3.87\pm0.02$ & $11.05\pm0.05$\\
  $1.5-2.0$ & $-3.78\pm0.03$ & $10.73\pm0.05$\\
  $2.0-2.5$ & $-3.72\pm0.04$ & $10.56\pm0.09$\\
  $2.5-3.0$ & $-3.65\pm0.05$ & $10.27\pm0.10$\\
  \hline  
\end{tabular}
\label{tab:sfmevo}
\end{center}
\end{table}
\end{center}

\section{RESULTS}
\subsection{Stellar Mass Growth}

\begin{figure}
\includegraphics[scale=0.5]{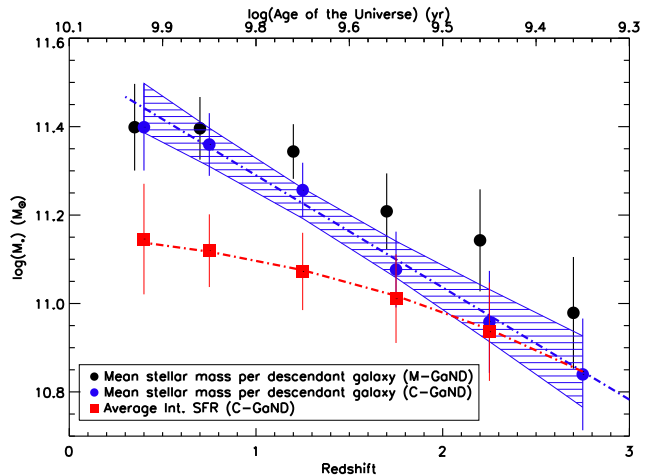}           
\caption{The mean stellar mass of galaxies per per $n=1\times10^{-4} \rm{Mpc}^{-3}$ descendant selected using the two number density selections as a function of redshift. The blue circles denote galaxies selected via the constant galaxy number density selection, and the black circles denote the major merger adjusted number density selected galaxies. This represents for the M\--GaND sample the total stellar mass that has already been created, but is in disparate objects. The blue dot dashed line shows the best simple linear fit to the C\--GaND data with the blue hashed region showing the 1 sigma uncertainty. The error bars are derived from Monte Carlo analyses incorporating the errors on stellar masses, redshift and number density. The red squares show the integrated SFR of the C\--GaND sample. This is calculated from the average galaxy SFR in each redshift bin and incorporates stellar mass loss due to stellar evolution derived from BC03 Chabrier model with sub-solar metallicity. The integrated SFRs are best fit by a power law shown in Equation 16.}
\label{massgrowth}
\end{figure}

Figure \ref{massgrowth} shows the evolving mean stellar mass per per $n=1\times10^{-4} \rm{Mpc}^{-3}$ descendant for both the C\--GaND and M\--GaND selected galaxies as a function of redshift and look back time. This represents for the M\--GaND sample the total stellar mass that has already been created, but is in disparate objects. Figure \ref{npro} shows the mean number of disparate objects at this redshift. The blue circles show the C\--GaND selected sample with $n=1\times10^{-4}\rm{Mpc^{-3}}$ and the black circles show the M\--GaND selected sample starting at $z=0.3$ with $n=1\times10^{-4}\rm{Mpc^{-3}}$. The blue dot dashed line shows the best simple linear fit to the C\--GaND data with the form:
\begin{equation}
 M_{*}(z)=11.56\pm0.13-(0.26\pm0.03)z 
\end{equation}
The hashed area denotes the 1 sigma errors on this fit. The fit to the C\--GaND implies that the direct progenitors of local massive galaxies with stellar masses of $\sim4\times10^{11}\rm{M_{\odot}}$ assembled $75\pm9\%$ of their stellar mass at $0.3<z<3.0$. This is consistent with stellar mass growth rates found in other number density studies (e.g., \citealt{Lundgren2014}, \citealt{Marchesini2014})

\subsection{Star formation history of massive galaxies from $z=3$ to $0.3$}

\begin{figure}
\includegraphics[scale=0.5]{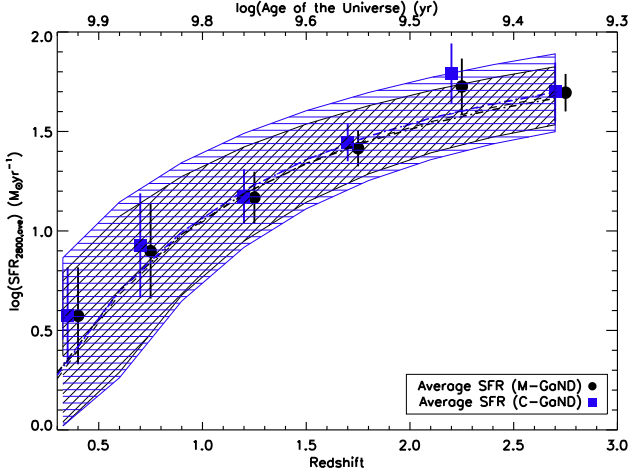}           
\caption{The average SFR of galaxies selected at a constant number density of $n=1\times10^{-4} \rm{Mpc}^{-3}$ (Blue squares) and galaxies selected using the major merger corrected number density as a function of redshift (Black circles). The SFRs are derived from the dust corrected UV luminosities. The average SFRs are fit with an exponentially declining model star formation history from $z=3.0$ to 0.3. The blue and black dotted lines show the best fits to each data set. The average SFRs are fit with an exponentially declining model star formation history from $z=3.0$ to $0.3$.}
\label{sfrhist}
\end{figure}

Using the average SFRs of the two galaxy populations we investigate the average star formation history of the massive galaxies over the range $0.3<z<3.0$. Figure \ref{sfrhist} shows the evolution of the dust corrected average SFR of the C\--GaND and M\--GaND galaxy populations. We observe that there is very little difference in the mean SFRs of the two samples, and there is a smooth decrease in the SFR from $z=3$ to 0.3. This decline can be fit by an exponentially declining model of the form:
 \begin{equation}
SFR(t)= SFR_{0} \times \rm{exp}(-t/\tau) 
\end{equation}
with $\tau =  2.3\pm0.6$ Gyr for the C\--GaND sample and $\tau= 2.3\pm0.6$ Gyr for the M\--GaND sample. This in contrast to the SFRs of massive galaxies at $z\gtrsim3$ which appear to be best fit with an increasing SFR model peaking at $z\simeq3.0$ (e.g. \citealt{Papovich2011}). We compare the star formation history for both galaxy samples to the star formation histories obtained for the same galaxies derived from SED fitting (see \S 2.3). We find that the average star formation history from SED fitting, $\tau_{SED} =  2.3\pm0.9$ Gyr, is very similar but with a larger error. 
We also examine how the star formation history of a population of galaxies varies as a function of the galaxy number density. 

We examine the star formation histories within a range of number densities from $n=5\times10^{-4} \rm{Mpc}^{-3}$ to $4\times10^{-5} \rm{Mpc}^{-3}$. We observe a slight change in the $\tau$ values within the number density selected samples. The C\--GaND selection has $\tau$ ranging from $2.4\pm0.5$ Gyr at $n=5\times10^{-4} \rm{Mpc}^{-3}$ to $2.2\pm0.5$ Gyr at $n =0.4\times10^{-4} \rm{Mpc}^{-3}$. The M\--GaND sample cannot be examined over the same range due to the galaxy sample dropping below the mass completeness limits at number densities lower than $n=1\times10^{-4} \rm{Mpc}^{-3}$. Therefore we examine it over a smaller range in number density from the studied $n=1\times10^{-4} \rm{Mpc}^{-3}$ to $0.4\times10^{-4} \rm{Mpc}^{-3}$. The value for $\tau$ obtained from the best fit to the SFRs at $n=0.4\times10^{-4} \rm{Mpc}^{-3}$, is $\tau = 2.3\pm0.6$ Gyr, showing the same trend as the C\--GaND sample. We also fit this relation excluding the point at $z=3.0$ as it appears that galaxies possibly depart from the exponentially declining model of SF at this redshift (\citealt{Papovich2011}). We find that even with excluding this redshift bin we recover essentially the same result.

From Section 2.5.2 if we assume zero dust correction for passive galaxies the star formation history for the $n=1\times10^{-4} \rm{Mpc}^{-3}$ C\--GaND sample changes to $\tau_{nodust}=1.7\pm0.7$ Gyr, within error of the full dust correction sample. This is also a hard lower limit on the star formation history due to the dust correction applied. 

Using the average SFRs of the C\--GaND sample we examine the stellar mass contribution of the SFR to the direct progenitors of massive galaxies over time. We study this directly by integrating the average SFRs from $0.3<z<3.0$ to obtain a total stellar mass added via SF. As the time scales involved within this integration are much larger than the main sequence lifetimes of high mass stars we need to consider the effect of the loss in stellar mass that will occur due to stellar evolution. 

To do this we used \cite{Bruzual2003} stellar population models with varying metallicity from $1/50$th solar, to solar, to estimate the fraction of the stellar mass created via SF that will be lost between integration steps. This fraction of the stellar mass is taken out of the integration. As an example, these models show that after $1$ Gyr of stellar evolution for a $1/2$ solar metallicity $\sim 35\%$ of the stellar mass produced at t$=0$ has been lost due to stellar evolution processes.

In the previous sections we examine the average total stellar mass growth of the selected massive galaxies populations seen in Figure \ref{massgrowth}. Also in Figure \ref{massgrowth} we plot the integrated SFR of the C\--GaND sample against redshift. From $z=3.0$ the integrated SFR is fitted using a power law of the form: 

\begin{equation}
{\rm log}(M_{\rm SFR}(z))=a-b*(1+z)^c 
\end{equation}

\noindent We find the best fit to all the free parameters for the $n=1\times10^{-4} \rm{Mpc}^{-3}$ C\--GaND sample is: $a=11.2\pm0.1, b=2\pm1\times10^{-2}$ and $c=3\pm1$. We find that between $1.5<z<3.0$ the stellar mass produced via the integrated SF can account for a large fraction, $\sim60\%$, of the total stellar mass growth over this redshift range. This implies that  SF is the dominant stellar mass growth process at these redshifts, and consequently the stellar mass growth from mergers must be smaller in comparison at $1.5<z<3.0$. 

At lower redshifts, $0.3<z<1.5$, the SF only accounts for $\sim0.1$ dex of stellar mass growth, wherein at the same redshift the total stellar mass grows by $\sim0.5$ dex. Using the results of this stellar mass build up in the C\--GaND sample we calculate the stellar mass added to the progenitor galaxies via all mergers across the redshift range $0.3<z<3.0$. The total mass deficit between the total stellar mass and the integrated SFR at $z=0.3$ is $\Delta M_{*} =(1.3\pm0.6)\times10^{11} \rm{M_{\odot}}$. As the integrated SFR at low redshift cannot account for the total stellar mass growth, mergers must be taking over as the dominant process of formation for the progenitors of local massive galaxies at $z=1.5$. In the next section we use these results, plus the results from the M\--GaND selected galaxies to calculate the stellar mass added via minor mergers.

\subsection{Galaxy Formation From Minor Mergers}
\begin{figure}
\includegraphics[scale=0.5]{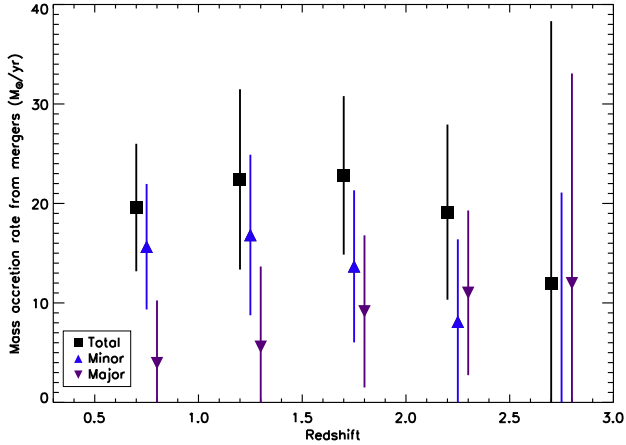}           
\caption{The total, minor and major merger accretion rate as a function of redshift in units of $\rm{M_{\odot}yr^{-1}}$. This is calculated from the deficit between the integrated SFR and the observed mass growth shown in Figure \ref{massgrowth}. The errors are calculated from Monte Carlo analyses incorporating the errors on the redshift, total stellar mass and the star formation rate. The black squares show the total merger rate, the blue upward pointing triangles show the minor merger rate  and the purple downward pointing triangles show the major merger rate.}
\label{mergerrate}
\end{figure}
\begin{figure}
\includegraphics[scale=0.5]{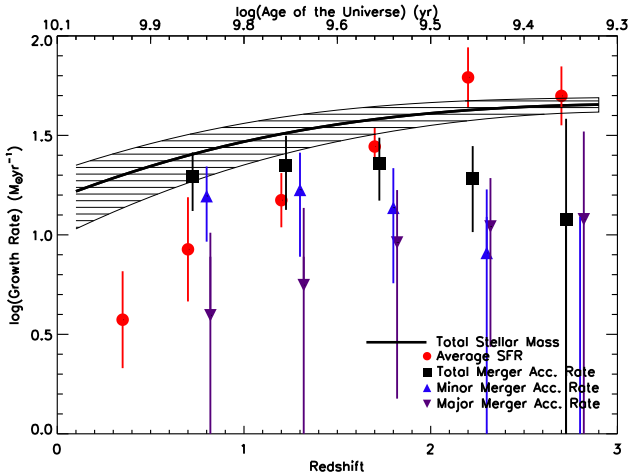}           
\caption{Growth rate of the number density selected galaxies as a function of redshift. The total growth rate is derived from the total stellar mass evolution shown in Figure \ref{massgrowth}. The black solid line shows the total stellar mass growth rate of the C\--GaND sample. The hashed region around the line show the 1 $\sigma$ uncertainty of the stellar mass growth rates derived from our Monte Carlo analysis. The red circles show the average SFR of the C\--GaND sample. The black squares show the calculated total merger rate for the C\--GaND sample. The blue upward pointing triangles show the minor merger rate and the purple downward pointing triangles show the major merger rate. See \S4.3 for full details on how these are derived. All error bars in this figure are derived from Monte Carlo analysis incorporating the errors of stellar masses, redshifts, selection criteria and SFRs.}
\label{growthrate}
\end{figure}
As discussed before in \S 1 the main two methods for increasing a galaxy's stellar mass are star formation and mergers. Therefore the growth of the stellar mass density ($\rho_{*}$) of a number density selected sample can be written as:
 \begin{equation}
\rho_{*}(z_{0})=\rho_{*}(z_{1})+\int_{z_{0}}^{z_{1}}\rho_{\rm{sfr}}(z)\,{\rm{d}}z+\int_{z_{0}}^{z_{1}}\rho_{\rm{m}} (z)\,{\rm{d}} z
\end{equation}
where $\rho_{*}(z_{0})$ and $\rho_{*}(z_{1})$ is the stellar mass density of the sample at different redshifts, where $z_{1}>z_{0}$, and $\rho_{SFR}(z)$ is the star formation rate density of the sample corrected for stellar evolution. This is integrated over the redshift range of interest to give a total stellar mass density added via star formation between $z_{0}-z_{1}$. The value $\rho_{m}(z)$ is the stellar mass of all galaxy mergers, both major and minor mergers, per unit volume of the sample, which can also be integrated over the redshift range to yield a total stellar mass density added via mergers. 

As we are selecting galaxies above a number density threshold, the total stellar mass density added via mergers cannot be due to mergers within the selected population. Within the M\--GaND selection the stellar mass of all major mergers that a likely to happen between $0.3<z<3.0$ are already contained within the sample. Therefore stellar mass density increase from the M\--GaND sample must be added from galaxies at higher number densities, or rather lower galaxy stellar mass (minor mergers). 

The three variables $\rho_{*}(z=0.3)$, $\rho_{*}(z=3.0)$ and $\rho_{sfr}(z)$ are known from the previous sections in this study (see \S4.1 and \S4.2 respectively).
From this we calculate, using a rearranged Equation 16, that the total stellar mass density added via mergers over the redshift range $z=0.3-3.0$ for the two samples are: 
  \begin{equation}
\int_{0.3}^{3.0}\rho_{\rm{m,C\--GaND}}(z)\,{\rm{d}}z=13.9\pm5.0\times10^{6}\, \rm{M_{\odot}Mpc^{-3}}
\end{equation}
 \begin{equation}
 \int_{0.3}^{3.0}\rho_{\rm{m,M\--GaND}}(z)\,{\rm{d}}z=10.2\pm2.3\times10^{6}\, \rm{M_{\odot}Mpc^{-3}}
 \end{equation}

\noindent The C\--GaND selection result gives the total stellar mass density added via all mergers, and M\--GaND selection result gives the total stellar mass density added via only minor mergers due to the selection encompassing all major merger progenitors. Therefore we can write these values as:
  \begin{equation}
\int_{0.3}^{3.0}\rho_{\rm{m,total}}(z)\,{\rm{d}}z=\int_{0.3}^{3.0}\rho_{\rm{m,C\--GaND}}(z)\,{\rm{d}}z
 \end{equation}
  \begin{equation}
\int_{0.3}^{3.0}\rho_{\rm{m,minor}}(z)\,{\rm{d}}z=\int_{0.3}^{3.0}\rho_{\rm{m,M\--GaND}}(z)\,{\rm{d}}z
 \end{equation}
 From these values we also calculate the total stellar mass density added via major mergers to the C\--GaND sample using the follow equation:
  \begin{equation}
\int\rho_{\rm{m,major}}(z)\,{\rm{d}}z=\int\rho_{\rm{m,total}}(z)\,{\rm{d}}z-\int\rho_{\rm{m,minor}}(z)\,{\rm{d}}z
 \end{equation}
  \begin{equation}
\int_{0.3}^{3.0}\rho_{\rm{m,major}}(z)\,{\rm{d}}z=3.7\pm2.2\times10^{6}\, \rm{M_{\odot}Mpc^{-3}}
 \end{equation}
If we assume that the total merger rate has been constant over this redshift range this equates to an average change in the stellar mass density due to major mergers of $\rho_{\rm{m,major}}=4.6\pm2.2\times10^{-4} \rm{M_{\odot}Mpc^{-3}yr^{-1}}$, and an average change in the stellar mass density due to minor merger of $\rho_{\rm{m,minor}}=12.9\pm1.9\times10^{-4} \rm{M_{\odot}Mpc^{-3}yr^{-1}}$ over $0.3<z<3.0$. Factoring in the number density of these objects implies that the total stellar mass accretion rate per galaxy from major mergers is $5\pm2\, \rm{M_{\odot}yr^{-1}}$ and the total stellar mass accretion rate per galaxy from minor mergers is $13\pm9\, \rm{M_{\odot}yr^{-1}}$. The large uncertainties on these results are due in the uncertainty on the minor merger rate at high redshifts. This can be improved by better knowledge of the major merger rates and stellar mass functions. However it is clear from observations that the major merger rate is not constant across this redshift range but it not yet clear from observations if the minor merger rate changes with redshift (e.g. \citealt{Bluck2012}). We also note that the definition in terms of stellar mass for what is classified as a major and a minor merger changes with redshift.

The results of \cite{Bluck2012}, \cite{Lopez2012}, \cite{Xu2012},  suggest that the average satellite in a major merger is 0.5 times the central galaxy stellar mass. Therefore, an alternative estimate for the expected increase in stellar mass density due to major mergers is approximately $1.5\times N_m \times \rho_{m,C\--GaND}$. when applying this method we obtain a stellar mass density increase due to major mergers is $5.6\pm4.2\times10^{6}\, \rm{M_{\odot}Mpc^{-3}}$, which is broadly consistent with method of choice for this work.

 \begin{figure*}
\subfloat[]{\includegraphics[scale=0.5]{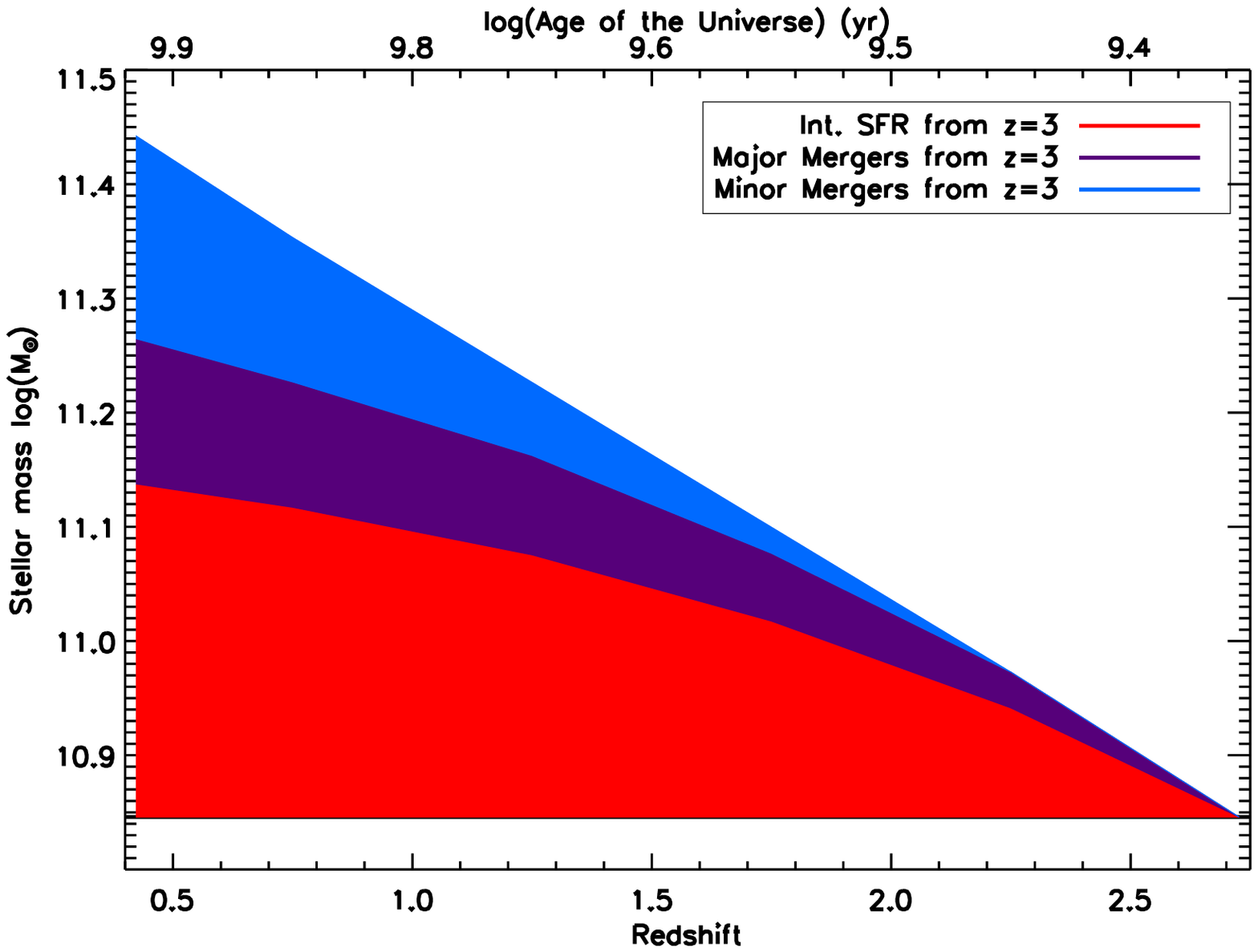}}
\subfloat[]{\includegraphics[scale=0.5]{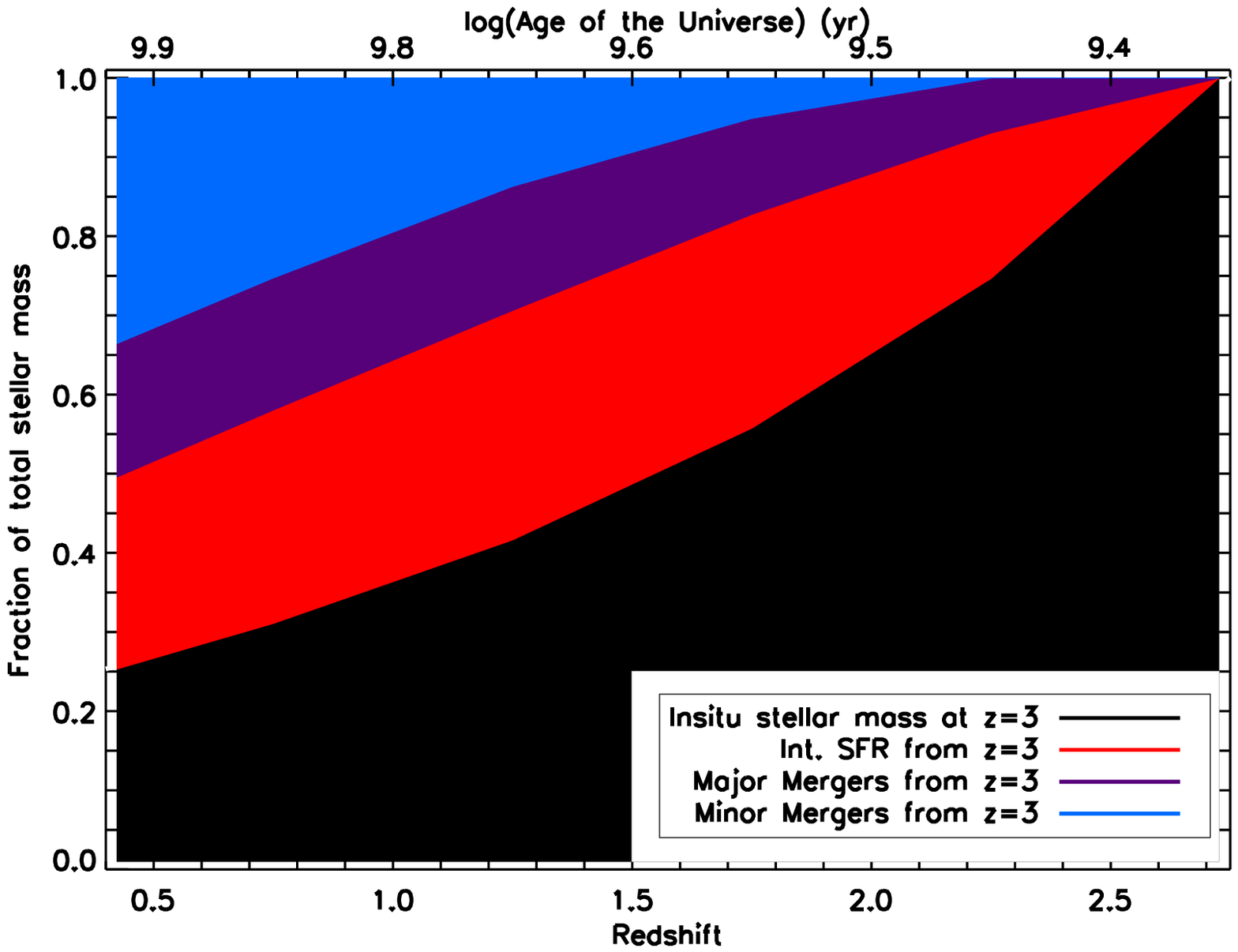}}
         
 \caption{The fraction of the total stellar mass created via SF since $z=3$ (red) and the stellar mass accreted from major mergers since $z=3$ (purple) and the stellar mass accreted from minor mergers since $z=3$ (blue) for the direct progenitors of local $\rm{log\, M_{*}} > 11.24$ massive galaxies corresponding to $\rm{log\, n}=-4.0$ (C\--GaND selected sample). (a) shows the total stellar mass growth and (b) shows the growth as a fraction of the total stellar mass at each redshift including the in-situ stellar mass at $z=3.0$ (black). Uncertainties on the fractions are shown in Figure \ref{massfracerrors}.}
 \label{massfrac}
 \end{figure*}

In the previous section we studied the difference in the integrated SFR and observed stellar mass growth of massive galaxies as a function of time. In this section we calculate the stellar mass deficit between the two relations and deduce the total stellar mass accreted over $0.3<z<3.0$ via mergers for the C\--GaND sample, $\Delta M_{*}=M_{m,total}=1.4\pm0.6\times10^{11} \rm{M_{\odot}}$. Therefore $50\pm20\%$ of the stellar mass of a massive galaxy at $z=0.3$ is accreted via merger accretion events since $z=3.0$. Dividing this figure into minor and major merger events, $34\pm14\%$ of the total stellar mass of a massive galaxy at $z=0.3$ is accreted from minor merger events and $17\pm15\%$ is accreted from major merger events. If we examine each redshift bin individually we can measure how the stellar mass accretion rate has changed due to various processes across the redshift range of this study. 

In Figure \ref{mergerrate} we show the calculated minor merger stellar mass accretion rate from the stellar mass density equations above applied to each redshift bin. Figure \ref{growthrate} shows the calculated minor merger rate compared to the SFR and stellar mass growth rate. As before the total merger rate is derived from the C\--GaND sample, and the minor merger rate from the M\--GaND sample. The two highest redshift bins have large uncertainties due to the SFR dominating at these redshifts. This does not rule out mergers at high redshift, but the effect caused via mergers must be small compared to the SFR at the same redshift. By examining the major mergers we find that the major merger accretion rate decreases towards lower redshifts. In Figure \ref{growthrate} we also find that the major merger rate in all of our redshift bins is lower than the SFR, therefore this implies that the major merger rate is at no point the dominant form of stellar mass growth between $0.3<z<3.0$. 

The minor merger rate however increases towards lower redshifts. In the highest redshift bins the minor merger rate is within the error consistent with zero but this again is due to the stellar mass added via the SFR being more significant at these times. Unlike the major merger rate in Figure \ref{growthrate} we see that the minor merger rate does become larger than the SFR at around $z=1.0$. Consequently the minor merger rate alone is the dominant form of stellar mass growth in the progenitors of local massive galaxies at $z<1$.

\subsection{Relative contributions to the stellar mass}

We compare the different stellar mass growth rates in massive galaxies for both selection criteria in Figure \ref{growthrate}. The total stellar mass growth rate for the C\--GaND sample is derived from the best fit to the total stellar mass growth shown in Figure \ref{massgrowth}. We see that the total stellar mass growth rate for massive galaxies has been declining since $z=3.0$. The blue points show the calculated minor merger rate as shown in Figure \ref{mergerrate}.

\begin{figure}
\includegraphics[scale=0.5]{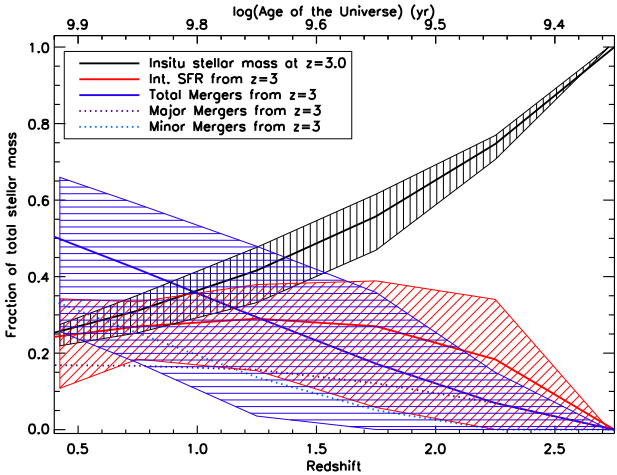}           
\caption{Errors on the fractional contributions to the total stellar mass derived from a Monte Carlo analysis. The stellar mass created via SF since $z=3.0$ (red), the total stellar mass accreted via all mergers since $z=3.0$ (blue) and the in-situ stellar mass at $z=3.0$ (black). The thin dotted lines show the fractional contribution of the major mergers (purple) and minor mergers (blue). }
\label{massfracerrors}
\end{figure}

We convert the values of the SFR, major and minor merger rates into the total amount of stellar mass created via these processes as a function of redshift shown in Figure \ref{massfrac}. In Figure \ref{massfrac} (a) we  see the contribution of the three processes to the total stellar mass growth since $z=3.0$. Figure \ref{massfrac} (b) shows the fractional contributions of in-situ stellar mass at $z=3.0$ (black), Integrated SFR (red), major mergers (purple) and minor mergers (blue) to the total stellar mass as a function of redshift. Figure \ref{massfracerrors} shows the errors on the fraction contributions derived from Monte Carlo analysis.

At our lowest redshift ($z=0.3$) the in-situ stellar mass at $z=3.0$ accounts for only $25\pm2\%$ of the total galaxy stellar mass. The stellar mass added via star formation accounts for $24\pm10\%$, and hence $51\pm20\%$ of the total galaxy stellar mass has been accreted via minor and major mergers. Therefore half of the stellar mass in local massive galaxies is not created within the galaxy, but has formed in other galaxies and has later been accreted. This is assuming that the cold gas that fuels the ongoing SFR originates from within the host progenitor galaxy, however this cold gas could also be accreted from the merger events or from the intergalactic medium, which we investigate in the next section. Within the mass obtained through mergers, $17\pm15\%$ of the total stellar mass has been accreted via major mergers, and the remaining $34\pm14\%$ via minor mergers. This implies that all three processes contribute approximately equal amounts of stellar mass to the total stellar mass of local massive galaxies from $z=3$ to $0.3$. Our work would seem to be in agreement with recent work by \cite{Lee2013} that showed, using merger tree simulations, that the most massive galaxies can obtain up to $70\%$ of their low redshift total stellar mass from mergers and accretion events. \cite{Dokkum2010} using a different constant number density technique than used in this paper show that $40\%$ of the total stellar mass of massive galaxies ($\rm{log}(M_{*})>11.45$) at z=0 was added through mergers and $10\%$ through star formation between $0<z<2$. Over the same redshift range this work finds that $\sim41\%$ of the total stellar mass of massive galaxies is added via all mergers and $\sim16\%$ is added via star formation. Conversely to the study, previous works (e.g. \citealt{Lopez2012} \citealt{Ferreras2013} \citealt{Ruiz2013}) have suggested that major mergers may play a more prominent role with up to $\sim60\%$ of a massive galaxies stellar mass growth at $z<2$ arising from major merger events.

If we assume that galaxies selected as passive via the UVJ selection technique have no dust correction to their SFRs (see \S 2.5) these results change slightly. The fraction of stellar mass created via star formation is on average $12\%$ smaller than the value above, within the errors quoted. Therefore the fraction of stellar mass accreted via all mergers increases to $62\pm15\%$ this breaks down to $41\pm10\%$ via minor mergers and $21\pm10\%$ via major mergers. The major merger fraction increases due to the objects within the M\--GaND sample being less affected by the change in dust correction. 

\subsection{Implications for gas accretion}

In this section we use our measured evolution in the total stellar mass, SFR and  mergers to predict the evolution in the total cold gas mass in the progenitors of local massive galaxies. We derive the cold gas mass surface density by using the global Schmidt\--Kennicutt relation calibrated for nearby star forming galaxies. The relation takes the form of:
\begin{equation}
\Sigma_{SFR}=1.7\pm0.5\times10^{-4}\left(\frac{\Sigma_{\rm{gas}}}{\rm{1\, M_{\odot}pc^{-2}}}\right)^{1.4\pm0.15}\, \rm{M_{\odot}yr^{-1}kpc^{-2}} 
\end{equation} 
where $\Sigma_{SFR}$ is the surface density of star formation, and $\Sigma_{\rm{gas}}$ is the surface density of cold gas (\citealt{Schmidt1959}, \citealt{Kennicutt1998b}). We calculate the star formation surface density for each galaxy based on the effective radius, $R_{e}$, obtained from \textsc{galfit} fitting S\'{e}rsic light profiles to the UDS K\--band images (see \S2.4). At high redshift \cite{O2012} showed that rest frame optical light profile is a good tracer for the profile of SF within massive galaxies. Using half of the measured SFR we obtain the gas mass surface density using Equation 23, and then calculate the total cold gas masses contained within these galaxies.

We can then express how the cold gas mass changes over time as:
\begin{equation}
M_{\rm{g}}(t)=M_{\rm{g}}(0)+M_{\rm{g,M}}(t)+M_{\rm{g,A}}(t)-\int SFR\,\rm{dt}+M_{\rm{g,recy.}}
\end{equation}
This is similar to \cite{C2013}, where we have an expression for the total gas mass of the galaxy at t=t, $M_{\rm{g}}(t)$, in terms of the total gas mass of the galaxy at t=0, $M_{\rm{g}}(0)$, the total gas mass accreted onto the galaxy via galaxy mergers from $t=t_{0}$ to $t=t_{f}$, $M_{\rm{g,M}}(t)$, the total amount of gas accreted onto the galaxy from the intergalactic medium i.e. with no corresponding increase in stellar mass from $t=t_{0}$ to $t=t_{f}$, $M_{\rm{g,A}}(t)$, as well as the amount of gas that is converted within the galaxy into stars, $-\int SFR\,\rm{dt}$, and the amount of stellar mass that is returned to the interstellar medium via stellar evolution, $M_{\rm{g,recy}}$.

\begin{figure}
\includegraphics[scale=0.5]{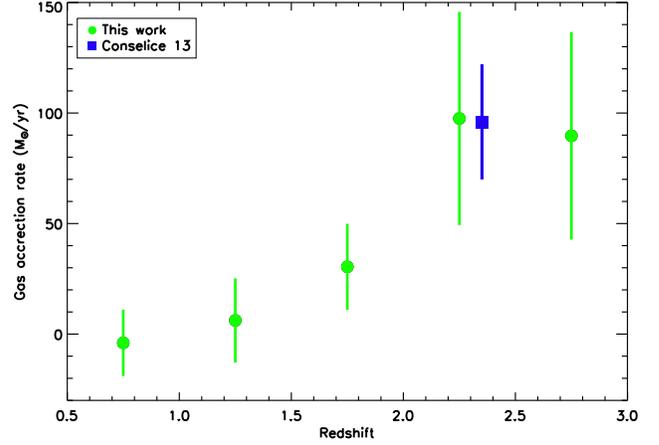}

 \caption{Cold gas accretion rate from the intergalactic medium of the C\--GaND galaxy sample. The green circles show the results of this work with the corresponding error bars denoting the 1 sigma error on the cold gas accretion rate derived from Monte Carlo methods. The blue square shows the average gas accretion rate as found by \protect \cite{C2013} over the redshift range $1.5<z<3.0$.}
 \label{gasfrac}
 \end{figure}

As we do not know the SFR of the galaxies that constitute the minor mergers we cannot calculate the exact total cold gas mass  added via minor mergers for these systems. Utilising other studies, \cite{C2013} calculated the average stellar mass to cold gas mass ratio of all galaxies from $M_* =10^{10.8} \rm{M_{\odot}}$ down to $M_*=10^{9.5} \rm{M_{\odot}}$  as  $f_{\rm{g}} = 1.03$. Using this information we calculate cold gas accretion needed across the redshift range $0.3<z<3.0$. We also know that cold gas can be ejected from the galaxy in winds from stellar or AGN sources. We account for the stellar outflows by assuming that the gas outflow rate is proportional to the SFR (e.g. \citealt{Erb2008},\citealt{Weiner2009}, \citealt{Bradshaw2013}). Therefore we add an extra term to equation 25 of $M_{\rm{g,outflow}}$ which we set equal to $\int SFR\,\rm{dt}$. Therefore we modify Equation 24 to account for this, and rearrange for $M_{\rm{g,A}}(t)$:
\begin{equation}
M_{\rm{g,A}}(t)=M_{\rm{g}}(t)-M_{\rm{g}}(0)-M_{\rm{g,M}}(t)+2\times\int SFR\,\rm{dt}-M_{\rm{g,recy}}
\end{equation}
\noindent Figure \ref{gasfrac} shows how the derived cold gas accretion rate changes with redshift. We see that the cold gas accretion rate has been in decline since $z=2.5$. At $z=2.5$ the progenitors of massive galaxies were accreting cold gas with an average rate of $97\pm49\, \rm{M_{\odot}yr^{-1}}$. From $z=2.0$ the cold gas accretion rate has undergone a decline to lower redshift ($z=0.3$). In fact at $z=0.3$ massive galaxies in the C\--GaND sample appear to have begun to have a negative gas accretion rate, $M_{\rm{g,A}}(z=0.3)=-4\pm15\, \rm{M_{\odot}yr^{-1}}$, This is likely due to other processes actively expelling gas from the host galaxy such as AGN. 

We compare this work with \cite{C2013} which also constrained the cold gas accretion rate within the redshift range of $1.5<z<3.0$. They found that within the redshift range of $1.5<z<3.0$ massive galaxies ($\rm{log}M_{*}>11.0 \rm{M_{\odot}}$) have an average cold gas accretion rate of $96\pm26\, \rm{M_{\odot}yr^{-1}}$. In the same redshift range we find that the progenitors of the local massive galaxies have an average cold gas accretion rate of $66\pm32\, \rm{M_{\odot}yr^{-1}}$. When we take into account the differences between the two works such as IMF and method of calculated SFR the two figures quoted are in agreement. We also examined different methods of calculating the cold gas outflow rate from massive galaxies (e.g. \citealt{Weiner2009}) and found that the cold gas accretion rate derived using these methods are within the error of the method used here.

\begin{center}
\begin{table}
\caption{Derived effective cold gas accretion rates from the intergalactic medium of the C\--GaND galaxy sample.}
\begin{center}
\begin{tabular}{| c | c | }
  \hline
  \hline
  $z$ & Accretion rate $M_{\odot}/yr$\\
  \hline             
  $0.5-1.0$  & $-4\pm15$\\
  $1.0-1.5$  & $6\pm19$\\
  $1.5-2.0$  & $30\pm19$\\
  $2.0-2.5$  & $97\pm48$\\
  $2.5-3.0$  & $89\pm47$\\
  \hline  
\end{tabular}
\label{tab:size}
\end{center}
\end{table}
\end{center}

\subsection{Size evolution}

\begin{center}
\begin{table}
\caption{C\--GaND average galaxy effective radius. Local ETG size derived from Shen et al. (2003) at the same stellar mass.}
\begin{center}
\begin{tabular}{| c | c | c | }
  \hline
  \hline
  $z$ & Average size (kpc)  & Local ETG size/Average size \\
  \hline             
  $0.3-0.5$ & $6.7\pm1.1$ & $1.2\pm_{-0.2}^{+0.2}$\\
  $0.5-1.0$ & $5.6\pm1.0$ & $1.4\pm_{-0.2}^{+0.3}$\\
  $1.0-1.5$ & $3.8\pm0.9$ & $1.8\pm_{-0.3}^{+0.5}$\\
  $1.5-2.0$ & $3.2\pm0.9$ & $1.7\pm_{-0.4}^{+0.7}$\\
  $2.0-2.5$ & $2.9\pm0.9$ & $1.6\pm_{-0.4}^{+0.8}$\\
  $2.5-3.0$ & $2.5\pm0.9$ & $1.6\pm_{-0.4}^{+0.9}$\\
  \hline  
\end{tabular}
\label{tab:size}
\end{center}
\end{table}
\end{center}

In this paper we investigate various properties of the progenitors of local massive galaxies using number density selection techniques. The sizes of galaxies, and how these have evolved over cosmic time is an area of galaxy evolution that we briefly examine using number density techniques.
  
Many papers examining the sizes of high redshift massive galaxies have found that on average their sizes are smaller, by a factor of between $2-4$ times, then present day galaxies of equal mass (e.g.\citealt{Daddi2005}, \citealt{Trujillo2007}, \citealt{Buitrago2008}, \citealt{Cimatti2008}, \citealt{Dokkum2008,Dokkum2010}, \citealt{Franx2008}, \citealt{Wel2008}, \citealt{Damjanov2009}, \citealt{Carrasco2010}, \citealt{Newman2010}, \citealt{Szomoru2011}, \citealt{Weinzirl2011}, \citealt{Lani2013}). This size evolution has been found to be most pronounced when linking high redshift passive massive galaxies to the passive massive galaxies in the local universe. This observed size evolution could be produced through various process such as AGN feedback (e.g. \citealt{Fan2008}), mergers (e.g. \citealt{Khochfar2006}), and star formation (e.g. \citealt{Dekel2009}, \citealt{O2012}). Therefore the observed size evolution is intrinsically linked to the growth of stellar mass. Another possible suggestion is that there is an inherent bias in the selection methods used in previous works that could enhance apparent observable size growth. It has been suggested that number density selection techniques could be a solution to this problem (e.g. \citealt{Poggianti2013}). For example \cite{Dokkum2010} investigated the size evolution within a constant number density selection over the range $0<z<2$, finding that the average galaxy size still increases by a factor of four.  

Most of these studies have examined size evolution using a cut in galaxy stellar mass in order to link galaxies across redshift. This method does not account for the stellar mass growth of galaxies that are below the stellar mass selection cut at high redshift. The number density selection techniques employed in this paper compensates for this, and can give us a cleaner sample of the progenitors of local massive galaxies. Using this sample of progenitor galaxies we can examine the size evolution in a more robust way.

Using the direct progenitor, C\--GaND, galaxy sample we investigate the evolution of the sizes of the progenitors of massive galaxies from $z=3.0$ to $z=0.3$. We do this by applying no passivity or morphological selection criteria to the sample and measure the size evolution of all the progenitor galaxies. As shown from this work a large fraction of the progenitors of local massive galaxies are highly star forming at high redshift and also appear to undergo a morphological change from disk\--like to spheroid\--like systems within the redshift range studied (\citealt{Buitrago2013}, \citealt{Mortlock2013}). 
 \begin{figure*}
\includegraphics[scale=1.0]{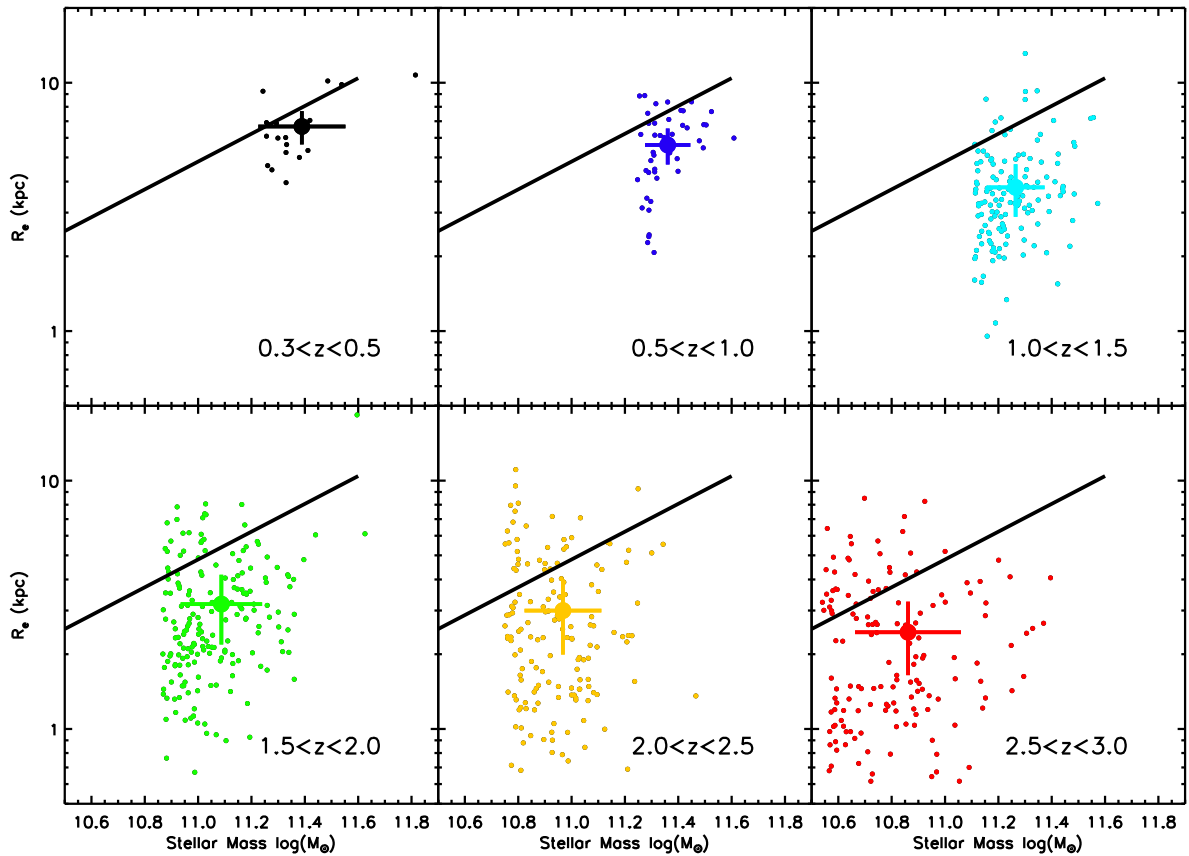}

 \caption{Galaxy size (effective radius) versus total stellar mass for the C\--GaND galaxy sample. The black line denotes the local early type galaxy relation modified from Shen et al. (2003). Within each redshift bin are plotted all the galaxies that reside within that redshift range (small circles) and the average stellar mass and size (large circle) with errors derived from Monte Carlo analysis within each redshift bin. }
 \label{size}
 \end{figure*}

Figure \ref{size} shows the effective radius versus total stellar mass of the whole C\--GaND galaxy sample split up into six redshift bins. In each bin we plot the galaxies that lie within the bin (small circles) and the average of the sample in both stellar mass and size (large circle with error bars). The solid back line denotes the local early type galaxy relation modified from \cite{Shen2003}. We compare the average galaxy size at each redshift to the local early type galaxy relation. We do this as the majority of the most massive galaxies lie on this relation in the local universe. When we compare the average points in each redshift bin to the local relation we find that all the  progenitor galaxies are smaller than equal mass early type galaxies in the local universe, ranging from a factor of 1.8 to 1.2 over the redshift range studied. 

Table \ref{tab:size} lists the average sizes of the progenitor galaxies and the ratio of the local size of an early type galaxy of the same stellar mass to the average size in each redshift bin. This would seem to be in disagreement with \cite{Dokkum2010}, however this could be due to differences between the selection techniques used. We find that the size evolution of a galaxy sample selected this way is on average slightly lower than the findings of other investigations into the size evolution of massive galaxies which have found that they grow in size by a factor of $2-4$ from redshift $z=3.0$ to the present day.

\section{Summary}
In this paper we investigate the role of star formation as well as major and minor mergers in relation to the total stellar mass growth of a constant number density selected galaxy sample within the redshift range of $0.3<z<3.0$. We use data from the UKIDSS UDS DR8, a deep near infra-red survey covering $\sim1$ square degree. We derive UV star formation rates for all the galaxies within this redshift range using SED fitted rest frame UV photometry accounting for dust and old stellar populations. 

We select the sample of massive galaxies using two number density methods; a constant number density selection (C\--GaND) and a major merger adjusted number density selection (M\--GaND). The major merger adjusted number density selection uses a selection that changes with time due to the rate of major mergers that occur over the redshift range studied. This selection traces the direct progenitor galaxies and the less massive galaxies that will merge with the direct progenitor galaxies at higher redshift. We use these selections to examine the average stellar mass growth of the progenitors of the most massive galaxies from $z=3.0$ to $z=0.3$ and disentangle the contributions of different processes of stellar mass growth. 

First we test the contamination of selecting progenitor galaxies using number density techniques using knowledge of the major merger rates and star formation histories. Contamination arises from lower mass galaxies entering the sample at lower redshifts via extreme star formation or high mass galaxies quenching and undergoing mergers. We find that the average contamination rate per redshift bin is $20-30\%$. We find that number density techniques are a factor of $3$ better at tracing progenitor than using a constant stellar mass selection technique. Our major results are:

\begin{itemize}
\item Local massive galaxies, with $\rm{log\,M_{*}}>11.24\, \rm{M_{\odot}}$, assemble $75\pm9\%$ of their $z=0.3$ total stellar mass between $0.3<z<3.0$.
\item Stellar mass created in star formation over the redshift range of $0.3<z<3.0$ comprises $24\pm8\%$ of the total stellar mass of massive galaxies at $z=0.3$. Examining the stellar mass contribution from total mergers between $0.3<z<3.0$ we find that the stellar mass added via mergers comprises $51\pm20\%$ of the total stellar mass of massive galaxies at $z=0.3$. We also find that the star formation history of the direct progenitors of the massive galaxies at $z=0.3$ can be defined by a declining $\tau$ model with $\tau=2.4\pm0.6\, \rm{Gyr^{-1}}$.
\item Star formation is the dominant process of stellar mass growth with the progenitor galaxies at $z>1.5$.
\item Total mergers (major and minor mergers combined) take over as the dominant process of stellar mass growth at $z<1.5$.
\end{itemize}
 Using the M\--GaND galaxy sample we separate the contributions of major and minor mergers to the total stellar mass growth. 
\begin{itemize}
\item We find that the minor merger rate of the progenitors of massive galaxies has been increasing with time since $z=3.0$ down to $z=0.3$.
\item Minor mergers become the dominant form of stellar mass growth in the progenitor galaxies at $z\leq1.0$.
\item The contribution from all minor mergers between $0.3<z<3.0$ is $34\pm14\%$ of the $z=0.3$ total galaxy stellar mass. All major mergers between $0.3<z<3.0$ contribute $17\pm15\%$ of the $z=0.3$ total galaxy stellar mass.
\item Major mergers are not the dominant form of stellar mass growth in the progenitor galaxies at any time between $0.3<z<3.0$.
\end{itemize}
Using the merger rate, SFR and stellar mass growth information we also investigate the cold gas accretion rate between $0.3<z<3.0$. We use the global Schimidt\--Kennicutt relation combined with work from \cite{C2013} to calculate the cold gas mass content of the progenitor galaxies at each redshift.
\begin{itemize}
\item We find that the cold gas accretion rate of the progenitor galaxies at $z=3.0$ is $97\pm49\, \rm{M_{\odot}yr^{-1}}$.
\item This cold gas accretion rate decreases with redshift until $z=0.3$.
\item The cold gas accretion rate in the lowest redshift bin is negative which is likely due to AGN feedback expelling the gas from a galaxy.
\end{itemize}
We also examine the size evolution of the constant number density selected sample using no passivity cuts and find that the sizes of the progenitors of massive galaxies range from a factor of $1.8$ to $1.2$ smaller than local early type galaxies of similar mass over the redshift range studied.

To further this work  large surveys such as the HSC survey and future telescopes such as JWST, E-ELT and Euclid will provide better constrained stellar mass functions that are required to explore these trends to a much higher precision.

\section*{Acknowledgements}
We thank the UDS team for their support and work on this survey and assistance on this paper. We acknowledge funding from the STFC and the Leverhulme trust for supporting this work. A.M. acknowledges funding from the a European Research Council Consolidator Grant (P.I. R. McLure)

\bibliographystyle{mnras}
\bibliography{bib}{}

\begin{thebibliography}{95}
\expandafter\ifx\csname natexlab\endcsname\relax\def\natexlab#1{#1}\fi

\bibitem[{Barden} et~al.(2012){Barden}, {H{\"a}u{\ss}ler}, {Peng}, {McIntosh}
  \& {Guo}]{Barden2012}
{Barden} M., {H{\"a}u{\ss}ler} B., {Peng} C.~Y., {McIntosh} D.~H., {Guo} Y.,
  2012, \mnras, 422, 449

\bibitem[{Bauer} et~al.(2011){Bauer}, {Conselice}, {P{\'e}rez-Gonz{\'a}lez}
  et~al.]{Bauer2011}
{Bauer} A.~E., {Conselice} C.~J., {P{\'e}rez-Gonz{\'a}lez} P.~G., et~al., 2011,
  \mnras, 417, 289

\bibitem[{Behroozi} et~al.(2013{\natexlab{a}}){Behroozi}, {Marchesini},
  {Wechsler}, {Muzzin}, {Papovich} \& {Stefanon}]{Behroozi2013b}
{Behroozi} P.~S., {Marchesini} D., {Wechsler} R.~H., {Muzzin} A., {Papovich}
  C., {Stefanon} M., 2013{\natexlab{a}}, \apjl, 777, L10

\bibitem[{Behroozi} et~al.(2013{\natexlab{b}}){Behroozi}, {Wechsler} \&
  {Conroy}]{Behroozi2013}
{Behroozi} P.~S., {Wechsler} R.~H., {Conroy} C., 2013{\natexlab{b}}, \apj, 770,
  57

\bibitem[{Bluck} et~al.(2009){Bluck}, {Conselice}, {Bouwens} et~al.]{Bluck2009}
{Bluck} A.~F.~L., {Conselice} C.~J., {Bouwens} R.~J., et~al., 2009, \mnras,
  394, L51

\bibitem[{Bluck} et~al.(2012){Bluck}, {Conselice}, {Buitrago}
  et~al.]{Bluck2012}
{Bluck} A.~F.~L., {Conselice} C.~J., {Buitrago} F., et~al., 2012, \apj, 747, 34

\bibitem[{Bourne} et~al.(2012){Bourne}, {Maddox}, {Dunne} et~al.]{Bourne2012}
{Bourne} N., {Maddox} S.~J., {Dunne} L., et~al., 2012, \mnras, 421, 3027

\bibitem[{Bradshaw} et~al.(2013){Bradshaw}, {Almaini}, {Hartley}
  et~al.]{Bradshaw2013}
{Bradshaw} E.~J., {Almaini} O., {Hartley} W.~G., et~al., 2013, \mnras, 433, 194

\bibitem[{Brammer} et~al.(2008){Brammer}, {van Dokkum} \& {Coppi}]{EAZY2008}
{Brammer} G.~B., {van Dokkum} P.~G., {Coppi} P., 2008, \apj, 686, 1503

\bibitem[{Bridge} et~al.(2010){Bridge}, {Carlberg} \& {Sullivan}]{Bridge2010}
{Bridge} C.~R., {Carlberg} R.~G., {Sullivan} M., 2010, \apj, 709, 1067

\bibitem[{Bruzual} \& {Charlot}(2003)]{Bruzual2003}
{Bruzual} G., {Charlot} S., 2003, \mnras, 344, 1000

\bibitem[{Buitrago} et~al.(2008){Buitrago}, {Trujillo}, {Conselice}, {Bouwens},
  {Dickinson} \& {Yan}]{Buitrago2008}
{Buitrago} F., {Trujillo} I., {Conselice} C.~J., {Bouwens} R.~J., {Dickinson}
  M., {Yan} H., 2008, \apjl, 687, L61

\bibitem[{Buitrago} et~al.(2013){Buitrago}, {Trujillo}, {Conselice} \&
  {H{\"a}u{\ss}ler}]{Buitrago2013}
{Buitrago} F., {Trujillo} I., {Conselice} C.~J., {H{\"a}u{\ss}ler} B., 2013,
  \mnras, 428, 1460

\bibitem[{Bundy} et~al.(2009){Bundy}, {Fukugita}, {Ellis}, {Targett}, {Belli}
  \& {Kodama}]{Bundy2009}
{Bundy} K., {Fukugita} M., {Ellis} R.~S., {Targett} T.~A., {Belli} S., {Kodama}
  T., 2009, \apj, 697, 1369

\bibitem[{Calzetti}(2001)]{Calzetti2001}
{Calzetti} D., 2001, PASP, 113, 1449

\bibitem[{Calzetti} et~al.(1994){Calzetti}, {Kinney} \&
  {Storchi-Bergmann}]{Calzetti1994}
{Calzetti} D., {Kinney} A.~L., {Storchi-Bergmann} T., 1994, \apj, 429, 582

\bibitem[{Carrasco} et~al.(2010){Carrasco}, {Conselice} \&
  {Trujillo}]{Carrasco2010}
{Carrasco} E.~R., {Conselice} C.~J., {Trujillo} I., 2010, \mnras, 405, 2253

\bibitem[{Charlot} \& {Fall}(2000)]{Charlot2000}
{Charlot} S., {Fall} S.~M., 2000, \apj, 539, 718

\bibitem[{Cimatti} et~al.(2008){Cimatti}, {Cassata}, {Pozzetti}
  et~al.]{Cimatti2008}
{Cimatti} A., {Cassata} P., {Pozzetti} L., et~al., 2008, \aap, 482, 21

\bibitem[{Cole} et~al.(2001){Cole}, {Norberg}, {Baugh} et~al.]{Cole2001}
{Cole} S., {Norberg} P., {Baugh} C.~M., et~al., 2001, \mnras, 326, 255

\bibitem[{Conselice}(2006)]{Conselice2006}
{Conselice} C.~J., 2006, \apj, 638, 686

\bibitem[{Conselice} et~al.(2013){Conselice}, {Mortlock}, {Bluck},
  {Gr{\"u}tzbauch} \& {Duncan}]{C2013}
{Conselice} C.~J., {Mortlock} A., {Bluck} A.~F.~L., {Gr{\"u}tzbauch} R.,
  {Duncan} K., 2013, \mnras, 430, 1051

\bibitem[{Daddi} et~al.(2007){Daddi}, {Dickinson}, {Morrison}
  et~al.]{Daddi2007}
{Daddi} E., {Dickinson} M., {Morrison} G., et~al., 2007, \apj, 670, 156

\bibitem[{Daddi} et~al.(2005){Daddi}, {Renzini}, {Pirzkal} et~al.]{Daddi2005}
{Daddi} E., {Renzini} A., {Pirzkal} N., et~al., 2005, \apj, 626, 680

\bibitem[{Damjanov} et~al.(2009){Damjanov}, {McCarthy}, {Abraham}
  et~al.]{Damjanov2009}
{Damjanov} I., {McCarthy} P.~J., {Abraham} R.~G., et~al., 2009, \apj, 695, 101

\bibitem[{de Ravel} et~al.(2011){de Ravel}, {Kampczyk}, {Le F{\`e}vre}
  et~al.]{Ravel2011}
{de Ravel} L., {Kampczyk} P., {Le F{\`e}vre} O., et~al., 2011, ArXiv:1104.5470

\bibitem[{Dekel} et~al.(2009){Dekel}, {Sari} \& {Ceverino}]{Dekel2009}
{Dekel} A., {Sari} R., {Ceverino} D., 2009, \apj, 703, 785

\bibitem[{Erb}(2008)]{Erb2008}
{Erb} D.~K., 2008, \apj, 674, 151

\bibitem[{Fan} et~al.(2008){Fan}, {Lapi}, {De Zotti} \& {Danese}]{Fan2008}
{Fan} L., {Lapi} A., {De Zotti} G., {Danese} L., 2008, \apjl, 689, L101

\bibitem[{Ferreras} et~al.(2013){Ferreras}, {Trujillo},
  {M{\'a}rmol-Queralt{\'o}} et~al.]{Ferreras2013}
{Ferreras} I., {Trujillo} I., {M{\'a}rmol-Queralt{\'o}} E., et~al., 2013,
  ArXiv:1312.5317

\bibitem[{Fischera} \& {Dopita}(2005)]{Fischera2005}
{Fischera} J., {Dopita} M., 2005, \apj, 619, 340

\bibitem[{Franx} et~al.(2008){Franx}, {van Dokkum}, {Schreiber}, {Wuyts},
  {Labb{\'e}} \& {Toft}]{Franx2008}
{Franx} M., {van Dokkum} P.~G., {Schreiber} N.~M.~F., {Wuyts} S., {Labb{\'e}}
  I., {Toft} S., 2008, \apj, 688, 770

\bibitem[{Furusawa} et~al.(2008){Furusawa}, {Kosugi}, {Akiyama}, {Takata},
  {Sekiguchi} \& {Furusawa}]{Furusawa2008}
{Furusawa} H., {Kosugi} G., {Akiyama} M., {Takata} T., {Sekiguchi} K.,
  {Furusawa} J., 2008, in { Panoramic Views of Galaxy Formation and
  Evolution\/}, edited by T.~{Kodama}, T.~{Yamada}, K.~{Aoki}, vol. 399 of {
  Astronomical Society of the Pacific Conference Series\/},  131

\bibitem[{Grogin} et~al.(2011){Grogin}, {Kocevski}, {Faber} et~al.]{Grogin2011}
{Grogin} N.~A., {Kocevski} D.~D., {Faber} S.~M., et~al., 2011, \apjs, 197, 35

\bibitem[{Hartley} et~al.(2013){Hartley}, {Almaini}, {Mortlock}
  et~al.]{Hartley2013}
{Hartley} W.~G., {Almaini} O., {Mortlock} A., et~al., 2013, \mnras, 431, 3045

\bibitem[{Hilton} et~al.(2012){Hilton}, {Conselice}, {Roseboom}
  et~al.]{Hilton2012}
{Hilton} M., {Conselice} C.~J., {Roseboom} I.~G., et~al., 2012, \mnras, 425,
  540

\bibitem[{Hopkins} \& {Beacom}(2006)]{Hopkins2006}
{Hopkins} A.~M., {Beacom} J.~F., 2006, \apj, 651, 142

\bibitem[{Ilbert} et~al.(2010){Ilbert}, {Salvato}, {Le Floc'h}
  et~al.]{Ilbert2010}
{Ilbert} O., {Salvato} M., {Le Floc'h} E., et~al., 2010, \apj, 709, 644

\bibitem[{Kennicutt}(1983)]{Kennicutt1983}
{Kennicutt} Jr. R.~C., 1983, \apj, 272, 54

\bibitem[{Kennicutt}(1998{\natexlab{a}})]{Kennicutt1998}
{Kennicutt} Jr. R.~C., 1998{\natexlab{a}}, \araa, 36, 189

\bibitem[{Kennicutt}(1998{\natexlab{b}})]{Kennicutt1998b}
{Kennicutt} Jr. R.~C., 1998{\natexlab{b}}, \apj, 498, 541

\bibitem[{Khochfar} \& {Silk}(2006)]{Khochfar2006}
{Khochfar} S., {Silk} J., 2006, \apjl, 648, L21

\bibitem[{Koekemoer} et~al.(2011){Koekemoer}, {Faber}, {Ferguson}
  et~al.]{Koekemoer2011}
{Koekemoer} A.~M., {Faber} S.~M., {Ferguson} H.~C., et~al., 2011, \apjs, 197,
  36

\bibitem[{Lacey} \& {Cole}(1993)]{Lacy1993}
{Lacey} C., {Cole} S., 1993, \mnras, 262, 627

\bibitem[{Lani} et~al.(2013){Lani}, {Almaini}, {Hartley} et~al.]{Lani2013}
{Lani} C., {Almaini} O., {Hartley} W.~G., et~al., 2013, \mnras, 435, 207

\bibitem[{Lawrence} et~al.(2007){Lawrence}, {Warren}, {Almaini}
  et~al.]{Lawrence2007}
{Lawrence} A., {Warren} S.~J., {Almaini} O., et~al., 2007, MNRAS, 379, 1599

\bibitem[{Lee} \& {Yi}(2013)]{Lee2013}
{Lee} J., {Yi} S.~K., 2013, \apj, 766, 38

\bibitem[{Leja} et~al.(2013){Leja}, {van Dokkum} \& {Franx}]{Leja2013}
{Leja} J., {van Dokkum} P., {Franx} M., 2013, \apj, 766, 33

\bibitem[{L{\'o}pez-Sanjuan} et~al.(2012){L{\'o}pez-Sanjuan}, {Le F{\`e}vre},
  {Ilbert} et~al.]{Lopez2012}
{L{\'o}pez-Sanjuan} C., {Le F{\`e}vre} O., {Ilbert} O., et~al., 2012, \aap,
  548, A7

\bibitem[{Lotz} et~al.(2008){Lotz}, {Jonsson}, {Cox} \& {Primack}]{Lotz2008}
{Lotz} J.~M., {Jonsson} P., {Cox} T.~J., {Primack} J.~R., 2008, \mnras, 391,
  1137

\bibitem[{Lundgren} et~al.(2014){Lundgren}, {van Dokkum}, {Franx}
  et~al.]{Lundgren2014}
{Lundgren} B.~F., {van Dokkum} P., {Franx} M., et~al., 2014, \apj, 780, 34

\bibitem[{Madau} et~al.(1996){Madau}, {Ferguson}, {Dickinson}, {Giavalisco},
  {Steidel} \& {Fruchter}]{Madau1996}
{Madau} P., {Ferguson} H.~C., {Dickinson} M.~E., {Giavalisco} M., {Steidel}
  C.~C., {Fruchter} A., 1996, \mnras, 283, 1388

\bibitem[{Magdis} et~al.(2010){Magdis}, {Elbaz}, {Hwang} et~al.]{Magdis2010}
{Magdis} G.~E., {Elbaz} D., {Hwang} H.~S., et~al., 2010, \apjl, 720, L185

\bibitem[{Marchesini} et~al.(2014){Marchesini}, {Muzzin}, {Stefanon}
  et~al.]{Marchesini2014}
{Marchesini} D., {Muzzin} A., {Stefanon} M., et~al., 2014, ArXiv:1402.0003

\bibitem[{McCarthy} et~al.(2004){McCarthy}, {Le Borgne}, {Crampton}
  et~al.]{McCarthy2004}
{McCarthy} P.~J., {Le Borgne} D., {Crampton} D., et~al., 2004, \apjl, 614, L9

\bibitem[{Meurer} et~al.(1999){Meurer}, {Heckman} \& {Calzetti}]{Meurer1999}
{Meurer} G.~R., {Heckman} T.~M., {Calzetti} D., 1999, \apj, 521, 64

\bibitem[{Mortlock} et~al.(2011){Mortlock}, {Conselice}, {Bluck}
  et~al.]{Mortlock2011}
{Mortlock} A., {Conselice} C.~J., {Bluck} A.~F.~L., et~al., 2011, \mnras, 413,
  2845

\bibitem[{Mortlock} et~al.(2013){Mortlock}, {Conselice}, {Hartley}
  et~al.]{Mortlock2013}
{Mortlock} A., {Conselice} C.~J., {Hartley} W.~G., et~al., 2013, \mnras, 433,
  1185

\bibitem[{Muzzin} et~al.(2013){Muzzin}, {Marchesini}, {Stefanon}
  et~al.]{Muzzin2013}
{Muzzin} A., {Marchesini} D., {Stefanon} M., et~al., 2013, \apj, 777, 18

\bibitem[{Newman} et~al.(2010){Newman}, {Ellis}, {Treu} \& {Bundy}]{Newman2010}
{Newman} A.~B., {Ellis} R.~S., {Treu} T., {Bundy} K., 2010, \apjl, 717, L103

\bibitem[{Noeske} et~al.(2007){Noeske}, {Weiner}, {Faber} et~al.]{Noeske2007}
{Noeske} K.~G., {Weiner} B.~J., {Faber} S.~M., et~al., 2007, \apjl, 660, L43

\bibitem[{Ownsworth} et~al.(2012){Ownsworth}, {Conselice}, {Mortlock},
  {Hartley} \& {Buitrago}]{O2012}
{Ownsworth} J.~R., {Conselice} C.~J., {Mortlock} A., {Hartley} W.~G.,
  {Buitrago} F., 2012, \mnras, 426, 764

\bibitem[{Pannella} et~al.(2009){Pannella}, {Gabasch}, {Goranova}
  et~al.]{Pannella2009}
{Pannella} M., {Gabasch} A., {Goranova} Y., et~al., 2009, \apj, 701, 787

\bibitem[{Papovich} et~al.(2011){Papovich}, {Finkelstein}, {Ferguson}, {Lotz}
  \& {Giavalisco}]{Papovich2011}
{Papovich} C., {Finkelstein} S.~L., {Ferguson} H.~C., {Lotz} J.~M.,
  {Giavalisco} M., 2011, \mnras, 412, 1123

\bibitem[{Patel} et~al.(2013){Patel}, {van Dokkum}, {Franx} et~al.]{Patel2013}
{Patel} S.~G., {van Dokkum} P.~G., {Franx} M., et~al., 2013, \apj, 766, 15

\bibitem[{P{\'e}rez-Gonz{\'a}lez} et~al.(2008){P{\'e}rez-Gonz{\'a}lez},
  {Rieke}, {Villar} et~al.]{PGon2008}
{P{\'e}rez-Gonz{\'a}lez} P.~G., {Rieke} G.~H., {Villar} V., et~al., 2008, \apj,
  675, 234

\bibitem[{Poggianti} et~al.(2013){Poggianti}, {Calvi}, {Bindoni}
  et~al.]{Poggianti2013}
{Poggianti} B.~M., {Calvi} R., {Bindoni} D., et~al., 2013, \apj, 762, 77

\bibitem[{Pozzetti} et~al.(2010){Pozzetti}, {Bolzonella}, {Zucca}
  et~al.]{Pozzetti2010}
{Pozzetti} L., {Bolzonella} M., {Zucca} E., et~al., 2010, \aap, 523, A13

\bibitem[{Prevot} et~al.(1984){Prevot}, {Lequeux}, {Prevot}, {Maurice} \&
  {Rocca-Volmerange}]{Prevot1984}
{Prevot} M.~L., {Lequeux} J., {Prevot} L., {Maurice} E., {Rocca-Volmerange} B.,
  1984, \aap, 132, 389

\bibitem[{Reddy} \& {Steidel}(2009)]{Reddy2009}
{Reddy} N.~A., {Steidel} C.~C., 2009, \apj, 692, 778

\bibitem[{Ruiz} et~al.(2013){Ruiz}, {Trujillo} \&
  {M{\'a}rmol-Queralt{\'o}}]{Ruiz2013}
{Ruiz} P., {Trujillo} I., {M{\'a}rmol-Queralt{\'o}} E., 2013, ArXiv:1312.4533

\bibitem[{Schmidt}(1959)]{Schmidt1959}
{Schmidt} M., 1959, \apj, 129, 243

\bibitem[{Sersic}(1968)]{sersic1968}
{Sersic} J.~L., 1968, {Atlas de galaxias australes}

\bibitem[{Shen} et~al.(2003){Shen}, {Mo}, {White} et~al.]{Shen2003}
{Shen} S., {Mo} H.~J., {White} S.~D.~M., et~al., 2003, \mnras, 343, 978

\bibitem[{Simpson} et~al.(2006){Simpson}, {Mart{\'{\i}}nez-Sansigre},
  {Rawlings} et~al.]{Simpson2006}
{Simpson} C., {Mart{\'{\i}}nez-Sansigre} A., {Rawlings} S., et~al., 2006,
  \mnras, 372, 741

\bibitem[{Springel} et~al.(2005){Springel}, {White}, {Jenkins}
  et~al.]{Springel2005}
{Springel} V., {White} S.~D.~M., {Jenkins} A., et~al., 2005, \nat, 435, 629

\bibitem[{Szomoru} et~al.(2011){Szomoru}, {Franx}, {Bouwens}
  et~al.]{Szomoru2011}
{Szomoru} D., {Franx} M., {Bouwens} R.~J., et~al., 2011, \apjl, 735, L22

\bibitem[{Tacconi} et~al.(2010){Tacconi}, {Genzel}, {Neri} et~al.]{Tacconi2010}
{Tacconi} L.~J., {Genzel} R., {Neri} R., et~al., 2010, \nat, 463, 781

\bibitem[{Tresse} et~al.(2007){Tresse}, {Ilbert}, {Zucca} et~al.]{Tresse2007}
{Tresse} L., {Ilbert} O., {Zucca} E., et~al., 2007, \aap, 472, 403

\bibitem[{Trujillo} et~al.(2007){Trujillo}, {Conselice}, {Bundy}, {Cooper},
  {Eisenhardt} \& {Ellis}]{Trujillo2007}
{Trujillo} I., {Conselice} C.~J., {Bundy} K., {Cooper} M.~C., {Eisenhardt} P.,
  {Ellis} R.~S., 2007, \mnras, 382, 109

\bibitem[{Ueda} et~al.(2008){Ueda}, {Watson}, {Stewart} et~al.]{Ueda2008}
{Ueda} Y., {Watson} M.~G., {Stewart} I.~M., et~al., 2008, \apjs, 179, 124

\bibitem[{van der Wel} et~al.(2012){van der Wel}, {Bell}, {H{\"a}ussler}
  et~al.]{Wel2012}
{van der Wel} A., {Bell} E.~F., {H{\"a}ussler} B., et~al., 2012, \apjs, 203, 24

\bibitem[{van der Wel} et~al.(2008){van der Wel}, {Holden}, {Zirm}
  et~al.]{Wel2008}
{van der Wel} A., {Holden} B.~P., {Zirm} A.~W., et~al., 2008, \apj, 688, 48

\bibitem[{van Dokkum} et~al.(2008){van Dokkum}, {Franx}, {Kriek}
  et~al.]{Dokkum2008}
{van Dokkum} P.~G., {Franx} M., {Kriek} M., et~al., 2008, \apjl, 677, L5

\bibitem[{van Dokkum} et~al.(2013){van Dokkum}, {Leja}, {Nelson}
  et~al.]{Dokkum2013}
{van Dokkum} P.~G., {Leja} J., {Nelson} E.~J., et~al., 2013, \apjl, 771, L35

\bibitem[{van Dokkum} et~al.(2010){van Dokkum}, {Whitaker}, {Brammer}
  et~al.]{Dokkum2010}
{van Dokkum} P.~G., {Whitaker} K.~E., {Brammer} G., et~al., 2010, \apj, 709,
  1018

\bibitem[{Wake} et~al.(2006){Wake}, {Nichol}, {Eisenstein} et~al.]{Wake2006}
{Wake} D.~A., {Nichol} R.~C., {Eisenstein} D.~J., et~al., 2006, \mnras, 372,
  537

\bibitem[{Weiner} et~al.(2009){Weiner}, {Coil}, {Prochaska} et~al.]{Weiner2009}
{Weiner} B.~J., {Coil} A.~L., {Prochaska} J.~X., et~al., 2009, \apj, 692, 187

\bibitem[{Weinzirl} et~al.(2011){Weinzirl}, {Jogee}, {Conselice}
  et~al.]{Weinzirl2011}
{Weinzirl} T., {Jogee} S., {Conselice} C.~J., et~al., 2011, \apj, 743, 87

\bibitem[{Whitaker} et~al.(2012){Whitaker}, {van Dokkum}, {Brammer} \&
  {Franx}]{Whitaker2012}
{Whitaker} K.~E., {van Dokkum} P.~G., {Brammer} G., {Franx} M., 2012, \apjl,
  754, L29

\bibitem[{Wijesinghe} et~al.(2012){Wijesinghe}, {Hopkins}, {Brough}
  et~al.]{Wijesinghe2012}
{Wijesinghe} D.~B., {Hopkins} A.~M., {Brough} S., et~al., 2012, \mnras, 423,
  3679

\bibitem[{Wijesinghe} et~al.(2010){Wijesinghe}, {Hopkins}, {Kelly}, {Welikala}
  \& {Connolly}]{Wijesinghe2010}
{Wijesinghe} D.~B., {Hopkins} A.~M., {Kelly} B.~C., {Welikala} N., {Connolly}
  A.~J., 2010, \mnras, 404, 2077

\bibitem[{Wilkins} et~al.(2008){Wilkins}, {Trentham} \& {Hopkins}]{Wilkins2008}
{Wilkins} S.~M., {Trentham} N., {Hopkins} A.~M., 2008, \mnras, 385, 687

\bibitem[{Williams} et~al.(2009){Williams}, {Quadri}, {Franx}, {van Dokkum} \&
  {Labb{\'e}}]{Williams2009}
{Williams} R.~J., {Quadri} R.~F., {Franx} M., {van Dokkum} P., {Labb{\'e}} I.,
  2009, \apj, 691, 1879

\bibitem[{Xu} et~al.(2012){Xu}, {Zhao}, {Scoville}, {Capak}, {Drory} \&
  {Gao}]{Xu2012}
{Xu} C.~K., {Zhao} Y., {Scoville} N., {Capak} P., {Drory} N., {Gao} Y., 2012,
  \apj, 747, 85

\end{thebibliography}

\label{lastpage}
\end{document}